\documentclass[12pt]{article}
\usepackage{epsfig, amsmath, amssymb}
\usepackage{graphicx,psfrag}
\usepackage{xcolor} 
\renewcommand{\baselinestretch}{1.22}

\usepackage{microtype}
\usepackage{kantlipsum}
\usepackage{float}

\newcommand{\nn}{\nonumber}
\newcommand{\be}{\begin{equation}}
\newcommand{\ee}{\end{equation}}
\newcommand{\ben}{\begin{equation}}
\newcommand{\een}{\end{equation}}
\newcommand{\bea}{\begin{eqnarray}}
\newcommand{\eea}{\end{eqnarray}}
\newcommand{\bA}{\begin{array}}
\newcommand{\eA}{\end{array}}
\newcommand{\bc}{\begin{center}}
\newcommand{\ec}{\end{center}}
\newcommand{\al}{\alpha}

\newcommand{\ra}{\rightarrow}

\newcommand{\ie}{{\it i.e.}}
\newcommand{\eg}{{\it e.g.}}

\begin{document}


\begin{titlepage}

\bc

\hfill 

\vspace{25mm}


{\Huge Small Schwarzschild de Sitter black holes, \\ [2mm]
the future boundary and islands}
\vspace{16mm}

{\large  Kaberi Goswami,\ \ K.~Narayan} \\
\vspace{3mm}
{\small \it Chennai Mathematical Institute, \\}
{\small \it SIPCOT IT Park, Siruseri 603103, India.\\}

\ec
\vspace{30mm}

\begin{abstract}
  We continue the study of 4-dimensional Schwarzschild de Sitter black
  holes in the regime where the black hole mass is small compared with
  the de Sitter scale, following arXiv:2207.10724 [hep-th]. The de
  Sitter temperature is very low compared with that of the black hole.
  We consider the future boundary as the location where the black hole
  Hawking radiation is collected. Using 2-dimensional tools, we find
  unbounded growth of the entanglement entropy of radiation as the
  radiation region approaches the entire future boundary.
  Self-consistently including appropriate late time islands emerging
  just inside the black hole horizon leads to a reasonable Page curve.
  We also discuss other potential island solutions which show
  inconsistencies.
\end{abstract}

\end{titlepage}

{\tiny 
\begin{tableofcontents}
\end{tableofcontents}
}


\section{Introduction}

The black hole information paradox \cite{Hawking:1976ra} has seen
fascinating progress over the last few years: In this context it is
perhaps best to regard this, not as a detailed understanding of black
hole microstates, but as the tension between the apparent unbounded
growth of entanglement entropy of Hawking radiation
\cite{Hawking:1975vcx} outside the black hole and the quantum
mechanics expectation that entanglement entropy must become small at
late times to recover purity of the original matter state
(see \eg\ \cite{Mathur:2009hf}, \cite{Almheiri:2012rt}, which review
various aspects of the information paradox). This
falling Page curve \cite{Page:1993wv}, \cite{Page:2013dx}, reflecting
the original purity, can be recovered when nontrivial, spatially
disconnected, island saddles for quantum extremal surfaces are
included \cite{Penington:2019npb}, \cite{Almheiri:2019psf},
\cite{Almheiri:2019hni}, \cite{Penington:2019kki},
\cite{Almheiri:2019qdq}.

Quantum extremal surfaces are extrema of the generalized gravitational
entropy \cite{Faulkner:2013ana,Engelhardt:2014gca} obtained from the
classical area of the entangling RT/HRT surface
\cite{Ryu:2006bv}-\cite{Rangamani:2016dms} after incorporating the
bulk entanglement entropy of matter.  Effective 2-dimensional models
allow explicit calculation, where 2-dim CFT techniques enable detailed
analysis of the bulk entanglement entropy.  The island, arising as a
nontrivial solution to extremization (near the black hole horizon, and
only at late times), reflects new replica wormhole saddles
\cite{Penington:2019kki,Almheiri:2019qdq} and serves to purify the
early Hawking radiation thereby lowering the entanglement entropy.
There is a large body of literature on various aspects of these
issues, reviewed in
\eg\ \cite{Almheiri:2020cfm,Raju:2020smc,Chen:2021lnq}: see
\eg\ \cite{Almheiri:2019yqk}-\cite{Franken:2023jas} 
for a partial list of investigations on black holes in various
theories, and also cosmological contexts. It is important to
note that several of these investigations are simply applications of
the island proposal, which appears to be self-consistent, even if it
cannot be rigorously derived in those contexts (see \eg\
\cite{Raju:2020smc} for an overall critical perspective, as well as
\cite{Laddha:2020kvp,Geng:2021hlu} and \cite{Bena:2022rna}).

\begin{figure}[H] 
\bc\includegraphics[width=40pc]{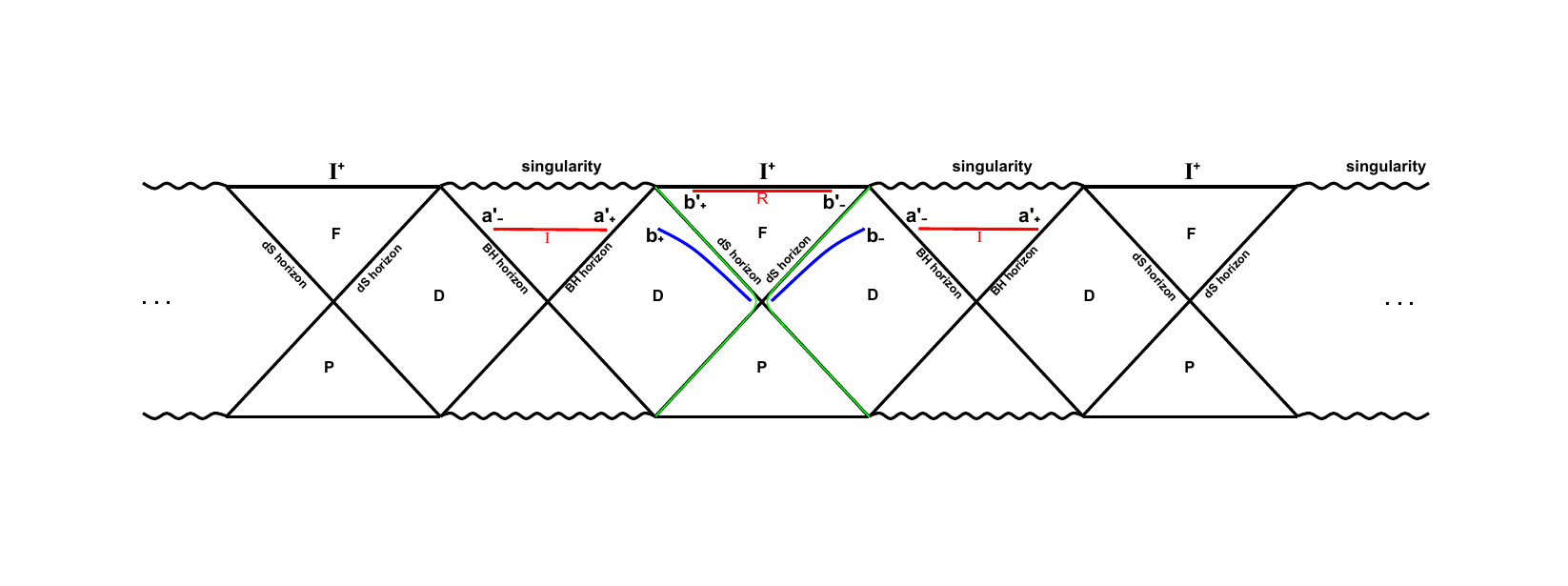}\ec \vspace{-18mm}
\caption{{ \label{figfuture boundary}
    \footnotesize{The Penrose diagram of a Schwarzschild de Sitter
      black hole, with radiation region near the future boundary
      $I^+$. Depicted are the radiation regions (blue lines) in the
      static patches, which are analytically continued to the
      radiation region $R \equiv [b'_+,b'_-]$ at $I^+$ and the late
      time island $I \equiv [a'_-,a'_+]$ on both sides.    
}}}
\end{figure}

This paper continues the study in \cite{Goswami:2022ylc} of ``small''
Schwarzschild de Sitter black holes, with the black hole mass $m$
small compared with the de Sitter scale $l$, but large enough that a
quasi-static approximation to the geometry is valid. The de Sitter
temperature is very low compared with that of the black hole so the
ambient de Sitter space is approximated as a frozen classical
background. For calculational purposes, we consider an effective
2-dim dilaton gravity model obtained by dimensional reduction, with
the bulk matter representing the black hole Hawking radiation
modelled as a 2-dim CFT propagating on this 2-dim space: this is
reasonable under the assumption that the s-wave Hawking modes are
dominant. We imagine that the black hole has formed from initial
matter in a pure state which is a reasonable approximation since the
de Sitter temperature is very low (more generally, the bulk matter CFT
is in a thermal state at the de Sitter temperature). In
\cite{Goswami:2022ylc}, we focussed on one black hole coordinate patch
in the Penrose diagram (roughly a line of alternating Schwarzschild
and de Sitter patches, see Figure~\ref{figfuture boundary}) and
considered observers in the static diamond patches far outside the
black hole but within the cosmological horizons. While the
entanglement entropy of the radiation region exhibits unbounded
growth, reflecting the information paradox for the black hole (which
has finite entropy), including appropriate island contributions
recovers finiteness of entanglement, and thereby expectations on the
Page curve. The island emerges at late times a little outside the
black hole horizon semiclassically.

\vspace{-13mm}

The Hawking radiation from the black hole is expected to cross the
cosmological horizon and eventually reach the future boundary where it
is collected (Figure~\ref{figfuture boundary}). In this paper, we
consider the point of view of these future boundary (meta)observers,
and look for semiclassical island resolutions of the black hole
information paradox with regard to a radiation region at the future
boundary. The future boundary is in a sense better defined (compared
to the static diamond) as a place where gravity is manifestly weak,
the space expanding indefinitely. The radiation region taken as an
interval with length labelled by $X$ (alongwith spheres) on the future
boundary can be parametrized via Kruskal coordinates $T, X$, defined
by analytic continuation from the static diamond coordinates. We find
that the entanglement entropy of Hawking radiation exhibits unbounded
growth in the spatial length $X$ along the future boundary,
inconsistent with the finiteness of black hole entropy, and reflecting
the information paradox. Using the island rule in the extremization of
the generalized entropy shows islands emerging for large values of $X$
a little inside the black hole horizon semiclassically: including the
island contributions recovers expectations on the Page curve.
This future boundary radiation region is entangled with
island regions around the horizons of black hole regions on both left
and right cosmological horizons (Figure \ref{figfuture boundary}):
this is expected since the future boundary receives Hawking modes from
both left and right black holes.  Our analysis has some parallels with
the island studies in \cite{Svesko:2022txo} for $dS_2$ arising under
reductions from Nariai limits of higher dim Schwarzschild de
Sitter. One might expect timelike separated quantum extremal surfaces
for the future boundary resulting in complex-valued entropies as are
known in pure de Sitter (see \cite{Chen:2020tes} for $dS_2$, and
\cite{Goswami:2021ksw} for reductions of higher dimensional Poincare
$dS$; see also \cite{Narayan:2015vda}, \cite{Sato:2015tta},
\cite{Narayan:2017xca}, \cite{Doi:2022iyj}, \cite{Narayan:2022afv} for
classical RT/HRT surfaces anchored at the future boundary). However
Schwarzschild de Sitter has a ``sufficiently wide'' Penrose diagram so
spacelike separated islands do exist here in accordance with physical
expectations for the black hole Page curve (thus we discard timelike
separated ones here).

In sec.~\ref{SdS:rev}, we review the Schwarzschild de Sitter geometry
and discuss parametrizations in various coordinate patches in
sec.~\ref{choice of coordinates}. Sec.~\ref{sec:noIsl} discusses the
entanglement entropy without islands (details in
App.~\ref{App:noIsland}), while sec.~\ref{sec:EEisland} discusses the
island calculation (details in App.~\ref{App:Island}).
App.~\ref{detailed calculation of previous paper} is a brief review of
the analysis in \cite{Goswami:2022ylc} for radiation entropy with
islands from the point of view of the static diamond observers.
App.~\ref{wrong1}-\ref{wrong2} discuss inconsistencies in other
potential island solutions, while App.~\ref{timelike separated qes}
discusses timelike separated quantum extremal surface solutions for
future boundary observers. Sec.~\ref{conclusions} contains a
Discussion on various aspects of our study.

\section{Small Schwarzschild de sitter black holes $\ra$\
  2-dim }\label{SdS:rev}

The Schwarzschild de Sitter black hole spacetime in $3+1$-dimensions
has the metric
\be\label{SdSst}
ds^2= -f(r)dt^2+\frac{dr^2}{f(r)}+r^2d\Omega_2^2\ ,  \qquad  
f(r)=1-\frac{2m}{r}-\frac{r^2}{l^2}\ .
\ee
This is a Schwarzschild black hole in de Sitter space
\cite{Gibbons:1977mu} with an ``outer'' cosmological (de Sitter)
horizon and an ``inner'' Schwarzschild horizon.
The general $d+1$-dimensional SdS spacetime is of similar form 
but with\ $f(r)=1-\frac{2m}{l} (\frac{l}{r})^{d-2}-\frac{r^2}{l^2}$,
and will have qualitative parallels. We are focussing on the 4-dim
case $SdS_4$ here: the function $f(r)$ is a cubic and the zeroes
of $f(r)$, \ie\ solutions to $f(r)=0$, give the horizon locations.
We parametrize this as
\bea\label{SdS4-fmrsrD}
f(r) = 1-\frac{2m}{r}-\frac{r^{2}}{l^{2}} = \frac{1}{l^2\,r}\,
(r_D-r)(r-r_S)(r+r_S+r_D)\,,\qquad\qquad\qquad \nn\\ [1mm]
r_Dr_S(r_D+r_S)=2ml^2\,,\qquad  r_D^2+r_Dr_S+r_S^2=l^2\ ;\qquad
0\leq r_S\leq r_D \leq l\ ;\qquad {m\over l}\leq {1\over 3\sqrt{3}}\ .
\eea
We will take the roots $r_S$ and $r_D$ to label the Schwarzschild
black hole and de Sitter (cosmological) horizons respectively.
(The third zero $-(r_{D}+r_{S})$ does not correspond to a physical
horizon.) The roots $r_S, r_D$ are constrained as above. 
The case with $m=0$, or $r_S=0,\ r_D=l$, is pure de Sitter space,
while the flat space Schwarzschild black hole has
$r_S=2m,\ r_D=l,\ l\ra\infty$.

The surface gravity at both horizons are generically distinct:
Euclidean continuations removing a conical singularity can be defined
at each horizon separately but not simultaneously at both
\cite{Ginsparg:1982rs} (see also
\cite{Bousso:1996au,Bousso:1997wi}). The only (degenerate) exception
is in an extremal, or Nariai, limit \cite{Nariai} where both
periodicities of Euclidean time match: the spacetime develops a nearly
$dS_2$ throat in this extremal limit \cite{Ginsparg:1982rs}.  More on
the nearly $dS_2$ limit and the wavefunction of the universe appears
in \cite{Maldacena:2019cbz}.  Related
discussions with some relevance to this paper also appears in
\cite{Fernandes:2019ige}.  In more detail, it can be seen that the
above horizon structure is valid for ${m\over l}<{1\over
  3\sqrt{3}}$\,, beyond which there are no horizons
\cite{Gibbons:1977mu}. The limit ${m\over l}={1\over 3\sqrt{3}}$ with
the cosmological and Schwarzschild horizon values coinciding, has
$r_S=r_D=r_0={l\over\sqrt{3}}$\ from (\ref{SdS4-fmrsrD}): this
extremal, or Nariai, limit has a near horizon $dS_2\times S^2$ throat.
Overall the range of physically interesting $r_S, r_D$ satisfies\ $0 <
r_S < r_0 < r_D$ for generic values.  The cosmological horizon is
``outside'' the Schwarzschild one since $r_S<r_D$.  The black hole
interior has $r<r_S$ with $r\ra 0$ the singularity.  The region
$r_D<r\leq\infty$ describes the future and past de Sitter universes,
with $r\ra\infty$ the future boundary $I^+$ (or past, $I^-$).  The
maximally extended Penrose diagram in Figure~\ref{figfuture boundary}
shows an infinitely repeating pattern of Schwarzschild coordinate
patches or ``unit-cells'' containing Schwarzschild black hole horizons
cloaking interior regions: these patches are bounded by cosmological
horizons on the left and right, with future/past universes beyond the
cosmological horizons.

As in \cite{Goswami:2022ylc}, we are considering the limit of a
``small'' black hole in de Sitter with
\be\label{m<<l}
m \ll l\,,\qquad\quad l\ra {\rm large}\qquad\Rightarrow\qquad r_D\gg r_S\ .
\ee
The horizon locations can then be found perturbatively to be\
$r_S\simeq 2m\,,\ r_D\simeq l-m \gg r_S$\,,\ from (\ref{SdS4-fmrsrD}).
This is a small black hole in a large accelerating universe, so the
ambient cosmology is effectively a frozen classical background while
the black hole Hawking evaporates. The black hole temperature is much
larger than the Gibbons-Hawking de Sitter temperature: from
\cite{Bousso:1996au,Bousso:1997wi} (see also
\cite{Shankaranarayanan:2003ya})
the surface gravities\ $\kappa_{BH, dS}={1\over 2\sqrt{f(ml^2)}}
\big\vert{df\over dr}\big\vert_{r_S,r_D}$ become\
${1\over 2\beta_{S,D}}\,{1\over\sqrt{1-3(m/l)^{2/3}}}$,\ with
$\beta_{S,D}$ in (\ref{beta}). Then the temperatures
$T={\kappa\over 2\pi}$ in the limit (\ref{m<<l}) become
\be\label{Tbh>>TdS}
T_{BH}\sim {1\over 8\pi m}\ ,\qquad T_{dS}\sim {1\over 2\pi l}\ ;\qquad\qquad
T_{dS} \ll T_{BH}\ .
\ee
The limit of asymptotically flat space is\
$r_D\sim l\ra \infty ,\ {r_D\over l} \ra 1,\ r_S\ra 2m$
and $T_{dS}\ra 0$ so the ambient de Sitter temperature vanishes.
Our discussions in this paper also pertain to these small
Schwarzschild de Sitter black holes.

\subsection{Coordinate parametrizations in various coordinate patches}\label{choice of coordinates}

We will describe various coordinate parametrizations in the various
coordinate choices in the Schwarzschild de Sitter spacetime, involving
Kruskal variables around the black hole horizon and around the
cosmological horizon.

In \cite{Goswami:2022ylc}, we considered the radiation region to be in
the static diamond bounded by the black hole and cosmological horizons
in the Schwarzschild de Sitter background: this static patch is
parametrized by certain Kruskal coordinates \cite{Guven:1990ubi} in
the vicinity of the black hole horizon.  For our present purposes, we
would like to analytically extend the Kruskal coordinates
$(U_{D},V_{D})$ defined in the static patch in the vicinity of the
cosmological horizon and $(U_{S},V_{S})$ near the black hole horizon
to a new set of Kruskal coordinates $(U_{D}',V_{D}')$ lying within the
future universe, near the future boundary, and $(U_{S}',V_{S}')$ in
the interior of black hole (inside the horizon) respectively.  We will
first define the set of coordinates in the static patch and then
analytically extend them beyond both horizons.

We will first recast the Schwarzschild de Sitter metric (\ref{SdSst})
in the static patch in terms of the Kruskal coordinates which are
regular at the cosmological horizon (but not in the vicinity of the
black hole horizon). Towards this, we define the tortoise coordinate
following \cite{Guven:1990ubi}:
\be\label{tortoise coordinate}
r^{\ast}=\int \frac{1}{f(r)} \,dr = \int \frac{1}{1-\frac{2m}{r}-\frac{r^{2}}{l^{2}}} \ dr=\int \frac{l^{2}r}{(r_{D}-r)(r-r_{S})(r+r_{S}+r_{D})} \ dr\ .
\ee
Taking $f(r)>0$ in the region $r_{S}<r<r_{D}$, this gives
\be\label{tortoise solution}
e^{r^{\ast}}=(r_{D}-r)^{-\beta_{D}}(r-r_{S})^{\beta_{s}}(r+r_{D}+r_{S})^{\beta_{M}}\ ,
\ee
with the parameters (which simplify $dr^\ast/dr$ to $1/f(r)$, and
satisfy $\beta_M+\beta_S=\beta_D$)
\be\label{beta}
\beta_{D}=\frac{l^{2}r_{D}}{(r_{D}-r_{S})(2r_{D}+r_{S})}\,,   \quad
\beta_{S}=\frac{l^{2}r_{S}}{(r_{D}-r_{S})(2r_{S}+r_{D})}\,,\quad
\beta_{M}=\frac{l^{2}(r_{D}+r_{S})}{(2r_{D}+r_{S})(2r_{S}+r_{D})}\,.
\ee
The $SdS_4$ metric (\ref{SdSst}) is recast 
as $ds^2= f(r)(-dt^2+{dr^\ast}^2)+r^{2}d\Omega_{2}^2$\,.\ We label the
spacetime coordinates in the left and right regions in the vicinity of
the cosmological horizon as
\be\label{b-b+beta}
b_+:\ \ \ (t,r)=(t_{b},b)\ ,\qquad
b_-:\ \ \ (t,r)=(-t_{b}+\frac{i\beta}{2},b)\ ;\qquad\quad
\beta=\frac{2\pi}{\alpha_{D}}\ .
\ee
This choice of $\beta$ is simply a convenient way of incorporating the
relative minus signs in the Kruskal coordinates in the left and right
regions through $e^{i\beta \al_D/2} = e^{i\pi} = -1$.
In the static patch, in the vicinity of the cosmological horizon, these
cosmological Kruskal coordinates $U_{D},V_{D}$, and the Schwarzschild de
Sitter metric become
\bea\label{SdS-Kruskal in SP}
b_{+}: \qquad U_{D_{+}} = e^{\alpha_{D}(t_{b}-r^{\ast})}\,,\quad V_{D_{+}} = -e^{-\alpha_{D}(t_{b}+r^{\ast})}\,,\qquad\qquad\qquad\qquad \nn \\ [2mm]
\alpha_{D}=\frac{1}{2\beta_{D}}\,;\quad
U_{D_{+}} V_{D_{+}} =  -e^{-2\al_D r^\ast}\ ,\quad \frac{U_{D_{+}}}{V_{D_{+}}} =  -e^{2\al_D t_b}\ ;
\quad\ 
ds^{2} = -\frac{dU_{D_+}dV_{D_+}}{{W_{b}}^2}+r^{2}d\Omega^{2}\ ,\nn\\ [2mm]
W_b = \sqrt{r}\,l\,\alpha_{D}(r_{D}-r)^{-\frac{(1-2\alpha_{D}\beta_{D})}{2}}(r-r_{S})^{-\frac{(1+2\alpha_{D}\beta_{S})}{2}}(r+r_{S}+r_{D})^{-\frac{(1+2\alpha_{D}\beta_{M})}{2}}\,.
\ \
\eea
The value of $\al_D$ here ensures regularity at the cosmological
horizon. ($\beta_M+\beta_S=\beta_D$ ensures that $W$ has dimensions of
inverse length.)\ With this parametrization of the left
and right time coordinates, we conveniently use the expressions in
(\ref{SdS-Kruskal in SP}), with $\beta$ automatically doing the left-right
book-keeping.

We are now considering an interval at the future boundary $I^+$ where
the Hawking radiation from the black hole is expected to be collected.
Towards parametrizing this future boundary radiation region, we will
analytically continue the cosmological Kruskal coordinates defined
above in the static patch to the region beyond the cosmological horizon
(\ie\ the future universe) keeping invariant, as usual, the metric
expressed in terms of the new cosmological Kruskal coordinates beyond
the cosmological horizon. Let us 
consider the analytic continuation in $(t_b,{r_b}^\ast)$ coordinates
as
\be\label{AC of b}
(t_b,{r_b}^\ast)\rightarrow
\big(X_{b'}=\alpha_D(t_b-\frac{i\pi}{2\alpha_D}),\,
T_{b'}=\alpha_D({r_b}^\ast-\frac{i\pi}{2\alpha_D})\big)\,.
\ee
Thus the new cosmological Kruskal coordinates at both ends $b'_+$ and
$b'_-$ of the future boundary radiation region are defined as
$(U'_{D_{+}},V'_{D_{+}})$ and $(U'_{D_{-}},V'_{D_{-}})$ respectively and the
Schwarzschild de Sitter metric in terms of $(X_{b'},T_{b'})$ coordinates
becomes
\bea\label{SdS-XT in FP}
b'_{+}: \qquad U'_{D_{+}} = e^{(X_{b'}-T_{b'})}\,,\qquad V'_{D_{+}} = e^{-(X_{b'}+T_{b'})} , \hspace{5cm}\nn \\ [1mm]
b'_{-}: \qquad U'_{D_{-}} = e^{-(X_{b'}+T_{b'})}\,,\qquad V'_{D_{-}} = e^{(X_{b'}-T_{b'})} , \hspace{5cm}\nn \\ [2mm]
U'_{D_{\pm}} V'_{D_{\pm}} =  e^{-2T_{b'}}\ ,\qquad \frac{U'_{D_{\pm}}}{V'_{D_{\pm}}} =  e^{\pm2X_{b'}}\ ;
\qquad ds^{2} = \frac{1}{\alpha_D^2} |f(r)|\, \big( -dT_{b'}^2+dX_{b'}^2 \big)
+ r^{2}d\Omega^{2} , \nn \\ [2mm]
\hspace{2.5cm}
|f(r)|= \frac{1}{l^2 r}(r-r_D)(r-r_S)(r+r_S+r_D)\ . \hspace{3cm}
\eea
The Schwarzschild de Sitter metric now becomes
\bea\label{SdS-kruskal in FP}
ds^{2} = -\frac{dU'_{D_{\pm}}dV'_{D_{\pm}}}{{W_{b'}}^2}+r^{2}d\Omega^{2}\ ,
\hspace{4cm}\nn\\ [2mm]
W_{b'} = \sqrt{r}\,l\,\alpha_{D}(r-r_{D})^{-\frac{(1-2\alpha_{D}\beta_{D})}{2}}(r-r_{S})^{-\frac{(1+2\alpha_{D}\beta_{S})}{2}}(r+r_{S}+r_{D})^{-\frac{(1+2\alpha_{D}\beta_{M}}{2})}\ .
\qquad
\eea

For our purposes, it is a reasonable approximation to look at the
s-wave sector of the black hole and consider the bulk matter as a
2-dim CFT: this enables the use of 2-dim CFT tools to study the
entanglement entropy of bulk matter. So, we will consider the same
dimensional reduction of the 4-dim Schwarzschild de Sitter spacetime
to a 2-dim background, as in \cite{Goswami:2022ylc}\ (see the general
reviews \cite{Strominger:1994tn,Grumiller:2002nm,Mertens:2022irh},
and \cite{Narayan:2020pyj} for related discussions, as well as
\cite{Bhattacharya:2020qil} for certain families of 2-dim
cosmologies).

Recalling from \cite{Goswami:2022ylc}, the reduction ansatz\
$ds^2_{(4)} = g^{(2)}_{\mu\nu} dx^\mu dx^\nu + \lambda^{-2} \phi\,d\Omega_2^2$
alongwith a Weyl transformation\ $g_{\mu\nu}=\phi^{1/2}\,g^{(2)}_{\mu\nu}$\
to absorb the dilaton kinetic term gives the 2-dim dilaton gravity
theory\ ${1\over 16\pi G_2}\int d^2x \sqrt{-g}\ (\phi {\cal R}
- {6\over l^2} \phi^{1/2} + 2\lambda^2\phi^{-1/2})$.
The lengthscale $\lambda^{-1}$ makes the dilaton $\phi$ dimensionless,
which then maps to the 4-dim transverse area of 2-spheres\
$4\pi\phi\over\lambda^2$\,. With $G_{_N}$ the 4-dim Newton constant,\
$G_2={G_{_N}\over V_2}$ and $V_2={4\pi\over\lambda^2}$\,, the 2-dim
theory has area term\
${\phi\over 4G_2}={4\pi\,r^2\over 4G_{_N}}$\ equivalent to the 4-dim
one. Our discussion is entirely gravitational so it is reasonable to
take the Planck length as the natural UV scale with\
$\lambda^{-1}\sim\epsilon_{UV}\sim l_P$\,.
So finally, the dilaton is $\phi=r^2\lambda^2$ and the 2-dim metric is 
\be\label{SdS-2d}
ds^{2} = - \lambda\,r\, \frac{dU_{D_{\pm}}'dV_{D_{\pm}}'}{{W_{b'}}^2}\, \equiv\,
- \frac{dU_{D_{\pm}}'dV_{D_{\pm}}'}{(W'_{b'})^2}\ ,
\ee
where $W'_{b'}={W_{b'}\over \sqrt{\lambda\,r}}= \frac{\alpha_D e^{-T_{b'}}}{\sqrt{\lambda\,r |f(r)|}}$ and $W_{b'}$ is the conformal factor given in (\ref{SdS-kruskal in FP}).

Next, we define a new set of Kruskal coordinates for the
location of the island boundary (the location of the
quantum extremal surface): this turns out be in the black hole interior
for the future boundary radiation region, so we require coordinate
parametrizations within the black hole horizon.
Towards this, we will again first recast the Schwarzschild de Sitter
metric (\ref{SdSst}) in the static patch in terms of Kruskal coordinates
regular at the black hole horizon. So we define the tortoise coordinate
$r^\ast$ in terms of the parameters $\beta_D, \beta_S$ and $\beta_M$ in
the same way as in (\ref{tortoise coordinate}), (\ref{tortoise solution})
and (\ref{beta}) in the static patch, in the vicinity of the black
hole horizon, with the $SdS_4$ metric recast as
$ds^2= f(r)(-dt^2+{dr^\ast}^2)+r^{2}d\Omega_{2}^2$\,. We label the
spacetime coordinates in the left and right regions in the vicinity of
the black hole horizon as
\be\label{a-a+beta}
a_+:\ \ \ (t,r)=(t_{a},a)\ ,\qquad
a_-:\ \ \ (t,r)=(-t_{a}+\frac{i\beta}{2},a)\ ;\qquad\quad
\beta=\frac{2\pi}{\alpha_{S}}\ .
\ee
Here also $\beta$ takes care of the relative minus signs in the
Kruskal coordinates in the left and right regions through
$e^{i\beta\al_S/2} = e^{i\pi} = -1$. In the static patch around the black
hole horizon, the Kruskal coordinates $U_{S},V_{S}$ and the Schwarzschild
de Sitter metric become
\bea\label{SdS-Kruskal in SP 1}
a_{+}: \qquad U_{S_{+}} = -e^{-\alpha_{S}(t_{a}-r^{\ast})}\,,\quad V_{S_{+}} = e^{\alpha_{S}(t_{a}+r^{\ast})} , \hspace{2cm}\nn \\ [2mm]
\alpha_{S}=\frac{1}{2\beta_{S}}\,;\quad
U_{S_{+}} V_{S_{+}} =  -e^{2\al_S r^\ast}\ ,\quad \frac{U_{S_{+}}}{V_{S_{+}}} =  -e^{-2\al_S t_a}\ ;
\quad 
ds^{2} = -\frac{dU_{S_+}dV_{S_+}}{{W_{a}}^2}+r^{2}d\Omega^{2}\ ,\nn\\ [2mm]
W_a = \sqrt{r}\,l\,\alpha_{S}(r_{D}-r)^{-\frac{(1+2\alpha_{S}\beta_{D})}{2}}(r-r_{S})^{\frac{(2\alpha_{S}\beta_{S}-1)}{2}}(r+r_{S}+r_{D})^{\frac{(2\alpha_{S}\beta_{M}-1}{2})}\ .
\qquad
\eea
The value of $\al_S$ here ensures regularity at the black hole horizon.
(noting $\beta_M+\beta_S=\beta_D$ we see that $W$ has dimensions of
inverse length.)\ With this parametrization of the left
and right time coordinates, we use the expressions in
(\ref{SdS-Kruskal in SP 1}) with $\beta$ doing the left-right
book-keeping.

Towards parametrizing the island boundary inside the black hole horizon,
we will analytically continue the spacetime coordinates defined in the
static patch near the black hole horizon keeping invariant the metric
in terms of the black hole interior Kruskal coordinates.
Let us consider the analytic continuation in $(t_a,{r_a}^\ast)$ coordinates
as
\be
(t_a,{r_a}^\ast)\rightarrow \big(X_{a'}=\alpha_S(t_a-\frac{i\pi}{2\alpha_S}),\,
T_{a'}=\alpha_S({r_a}^{\ast}+\frac{i\pi}{2\alpha_S})\big)\,.
\ee
Thus the new set of Kruskal coordinates at both the island boundaries
$a'_+$ and $a'_-$ are defined as $(U'_{S_{+}},V'_{S_{+}})$ and
$(U'_{S_{-}},V'_{S_{-}})$ respectively and the Schwarzschild de Sitter
metric in terms of $(X_{a'},T_{a'})$ coordinates becomes
\bea\label{SdS-XT in BHP}
a'_{+}: \qquad U'_{S_{+}} = e^{-(X_{a'}-T_{a'})}\,,\qquad V'_{S_{+}} = e^{(X_{a'}+T_{a'})} , \hspace{5cm}\nn \\ [1mm]
a'_{-}: \qquad U'_{S_{-}} = e^{(X_{a'}+T_{a'})}\,,\qquad V'_{S_{-}} = e^{-(X_{a'}-T_{a'})} , \hspace{5cm}\nn \\ [2mm]
U'_{S_{\pm}} V'_{S_{\pm}} =  e^{2T_{a'}}\ ,\qquad \frac{U'_{S_{\pm}}}{V'_{S_{\pm}}} =  e^{\mp2X_{a'}}\ ;
\qquad ds^{2} = \frac{1}{\alpha_S^2} |f(r)|\, \big(-dT_{a'}^2+dX_{a'}^2\big) + r^{2}d\Omega^{2} , \nn \\ [2mm]
\hspace{2cm} |f(r)|= \frac{1}{l^2 r}(r_D-r)(r_S-r)(r+r_S+r_D)\ .
\hspace{3cm}
\eea
The Schwarzschild de Sitter metric in terms of Kruskal coordinates becomes
\bea\label{SdS-kruskal in BHP}
ds^{2} = -\frac{dU'_{S_{\pm}}dV'_{S_{\pm}}}{{W_{a'}}^2}+r^{2}d\Omega^{2}\ ,   \hspace{4cm}\nn\\ [2mm]
W_{a'} = \sqrt{r}\,l\,\alpha_{S}(r_{D}-r)^{-\frac{(1+2\alpha_{S}\beta_{D})}{2}}(r_{S}-r)^{\frac{(2\alpha_{S}\beta_{S}-1)}{2}}(r+r_{S}+r_{D})^{\frac{(2\alpha_{S}\beta_{M}-1}{2})}\ .
\qquad
\eea
Here also after the same dimensional reduction, as in \cite{Goswami:2022ylc}, the 2-dim metric beyond the black hole horizon becomes
\be\label{SdS-2d 1}
ds^{2} = - \lambda\,r \frac{dU_{S_{\pm}}'dV_{S_{\pm}}'}{{W_{a'}}^2}\, \equiv\,
- \frac{dU_{S_{\pm}}'dV_{S_{\pm}}'}{(W'_{a'})^2}\ ,
\ee
where $W'_{a'}={W_{a'}\over \sqrt{\lambda\,r}}= \frac{\alpha_S e^{T_{a'}}}{\sqrt{\lambda\,r |f(r)|}}$ and $W_{a'}$ is the conformal factor
given in (\ref{SdS-kruskal in BHP}).

\section{Entanglement entropy: no island}\label{sec:noIsl}

In this section, we will evaluate the entanglement entropy of the
radiation at late times in the absence of any island. Here we have the radiation region $R$ within the interval $b'_{+}$ and $b'_{-}$ as shown in Figure \ref{figfuture boundary}. We have chosen the bulk matter to be within the above stated interval in some fixed $T$ slice near the future boundary. The entropy of the Hawking radiation is
\be\label{without island formula}
S_{matter} = S(R)\ .
\ee
We calculate the bulk entropy technically using 2-dimensional techniques where we
approximate the bulk matter by a 2-dim CFT propagating in the 2-dim
background. In the 2-dim CFT, the matter entanglement entropy for a single
interval $A=[x,y]$ is obtained from the replica formulation
\cite{Calabrese:2004eu,Calabrese:2009qy} 
after also incorporating in $d[x,y]$ the
conformal transformation to a curved space \cite{Almheiri:2019psf},
stemming from the $W'$-factor in the 2-dim metric (\ref{SdS-2d}),
\be\label{single interval}
S_{A}=\frac{c}{3}\log{d[x,y]} \,=\,
{c\over 6} \log \left({-\Delta U_S \Delta V_S\over W'|_x\,W'|_y}\right)\ .
\ee
So, we obtain the entropy of the bulk matter CFT of the radiation region as
\be\label{S_matter}
    S_{matter} = \frac{c}{3}\,\log\,[d(b_{+}',b_{-}')]\ .
\ee
Then we evaluate the bulk matter entropy near the future boundary in the
Schwarzschild de Sitter geometry (\ref{SdS-2d}) to obtain\
(suppressing $1/\epsilon_{UV}^2$ inside the logarithm, $\epsilon_{UV}$
the UV cutoff)
\be\label{matter entropy}
S_{\text{matter}} = \frac{c}{6} \, \log \left(\frac{(U_{D_{-}}'-U_{D_{+}}')(V_{D_{+}}'-V_{D_{-}}')}{W'_{b'_{+}}W'_{b'_{-}}}\right)\ .
\ee
Using the Kruskal coordinates (\ref{SdS-XT in FP}), the bulk matter
entanglement entropy then becomes
\be\label{Total entropy}
S = \frac{c}{6}\,\log \left[16 \lambda\beta_{D}^{2}\,(b'-r_{D})\,
\frac{(b'-r_{S})(b'+r_{S}+r_{D})}{l^2} \, \sinh^{2}(X_{b'})\right]\,.
\ee
Alongwith $\epsilon_{UV}^2$ and the discussions around (\ref{SdS-2d}),
it can be seen that the logarithm argument is dimensionless (noting
from (\ref{beta}) that $\beta_D$ has dimensions of length).
The details of the calculation are shown in Appendix \ref{App:noIsland}.
The late time approximation is done by considering large $X_{b'}$,
which means we are considering the entire constant $T$ slice: the
above result then approximates as

\be\label{approx.entropy}
S\approx const + \frac{c}{3}\,X_{b'}\ .
\ee
This linear growth of the bulk matter entropy with length $X_{b'}$
means that the entropy of the radiation will eventually be infinitely
larger than the Bekenstein-Hawking entropy of the black hole for large
$X_{b'}$. This inconsistency is the reflection of the black hole
information paradox from the future boundary point of view. See
\cite{Svesko:2022txo} for similar observations in $dS_2$.

To gain some intuition for this linear growth with ``length'' at the
future boundary relative to linear growth in ordinary time, it is
useful to compare the present situation with \cite{Goswami:2022ylc}
where we studied the evolution of the entanglement entropy of
radiation collected by observers labelled by $b_\pm$ in the left/right
static diamond patches (see Fig.~\ref{figfuture boundary}) at late
times \ie\ for large $|t_b|$.  In our present context, the future
boundary radiation region $(b'_+,b'_-)$ is defined by spacetime
coordinates $(X_b',T_b')$ obtained by analytic continuation (\ref{AC of b})
of the spacetime coordinates $(t_b,{r_b}^\ast)$ defined in the static
diamond patches.  Geometrically, using Fig.~\ref{figfuture boundary},
we see that points in the left and right static diamonds can be mapped
to points near the future boundary $I^+$ by drawing out light rays
from $b_+$ to $b'_+$ and $b_-$ to $b'_-$. In the late limit with
$|t_b|$ large, the points $b_\pm$ in the left/right static diamonds
(left/right ends of the blue radiation regions) move towards the top
end of the green lines (observer worldlines just inside the
cosmological horizon).  The corresponding lightrays map this to points
near the left/right ends of the future boundary, giving large lengths
$|X_{b'}|$, consistent with the analytic continuation (\ref{AC of b}).
In other words, the points $b'_\pm$ approach the ends of the future
boundary.  This is consistent with the picture of Hawking radiation
from the black hole eventually crossing the cosmological horizons and
reaching the future boundary, so that late times for static patch
observers map to large lengths for future boundary (meta)observers.
Our future boundary (meta)observers perspective here is reminiscent of
the ``census-taker'' who looks back into the past and collect data
\cite{Susskind:2007pv}: it would be fascinating to make this precise
and develop further.

\section{Late time entanglement entropy with island}\label{sec:EEisland}

The Hawking radiation from the black hole will eventually cross the
cosmological horizon and reach the future boundary (see
Figure~\ref{figfuture boundary}), where we imagine it is collected by
appropriate (meta)observers.
In this section, we will evaluate the entanglement entropy of the bulk
matter near the future boundary after including appropriate islands.
The island proposal \cite{Almheiri:2019hni} for the fine-grained entropy
of the Hawking radiation is 
\be\label{Island formula}
S(R)=min\left\{{ext\Big[\frac{Area(\partial I)}{4G_{_N}}
      + S_{matter}(R\cup I)\Big]}\right\}
\ee
where $R$ is the region far from the black hole where the radiation is
collected by distant observers and $I$ is a spatially disconnected
island around the horizon that is entangled with $R$. The intuition
here is that after about half the black hole has evaporated, the
outgoing Hawking radiation (roughly $I$) begins to purify the
early radiation (roughly $R$). This purification by the late Hawking
radiation of the early radiation reflects the entanglement between
the two parts, stemming from the picture of Hawking radiation as
production of entangled particle pairs near the horizon (taken as
vacuum). Thus $R\cup I$ purifies over time, its entanglement
decreasing. The decreasing area of the slowly evaporating
(approximately quasistatic) black hole then leads to $S(R)$
decreasing in time, recovering the falling Page curve expected
from unitarity of the original approximately pure state.

In the current case, the future boundary receives Hawking radiation
from both the left and right black hole horizons so we expect islands
on both left and right. Each island almost entirely covers the
corresponding black hole interior: the island boundaries are at
$a'_{+}$ and $a'_{-}$\ (Figure~\ref{figfuture boundary}).
The islands in question turn out to emerge just inside the black
hole horizon, so
\be\label{b'-rD>>rs-a', b>>rs}
b'-r_{D}\ \gg \ r_{S}-a'\ \sim\ 0\ .
\ee
Including an island $I$ at late times \ie\ for large $X_{b'}$
and $X_{a'}$, the effective radiation region becomes
$\Sigma_{rad} \cup I$. Now we make the assumption that the global vacuum
state is approximately pure: this is not strictly true since the bulk
matter CFT is expected to be at finite $dS$ temperature in the ambient
de Sitter space. However in the limit of a small mass black hole in
a very large $dS$ space with correspondingly very low $dS$
temperature, one can take the bulk matter to be at nearly zero
temperature and correspondingly in a global pure state. With this
assumption, one instead computes the entanglement entropy of the
complementary region $(\Sigma_{rad} \cup I)^{c}$, which comprises
the two intervals $[a'_{+},b'_{+}]$ and $[b'_{-},a'_{-}]$, which turns
out to be self-consistent.


The entanglement entropy for multiple disjoint intervals
\be
A=[x_{1},y_{1}]\ \cup\ [x_{2},y_{2}]\ \cup\ \ ....\ \ \cup\ [x_{N},y_{N}]
\ee
is more complicated, arising from the multi-point correlation
functions of twist operators: so it depends on not just the
central charge but detailed CFT information. In the limit where the
intervals are well-separated,  expanding the twist operator products
yields \cite{Calabrese:2004eu,Calabrese:2009qy,
  Calabrese:2009ez,Calabrese:2010he}
\be\label{multiple interval}
S_A = \frac{c}{3}\,\log\, { \prod_{i,j} d[x_{j},y_{i}] \over \,
  \prod_{i<j} d[x_{j},x_{i}]\ \ \prod_{i<j} d[y_{j},y_{i}]\, }
\ee
For two intervals $[x_1,y_1]\cup [x_2,y_2]$, this is a limit where
the cross ratio $x$ is small, \ie\ $x\ll 1$, with\
$x = {\, d[x_1,y_1]\, d[x_2,y_2]\,\over\, d[x_1,x_2]\,d[y_1,y_2]\,}$,\
and we use the Kruskal distances in (\ref{single interval}) in
constructing the cross-ratio.
In 2-dim CFTs with a holographic dual, this is the situation where the
two intervals $A, B$ are well-separated and their mutual information 
exhibits a disentangling transition \cite{Headrick:2010zt} with
$I[A,B]=S[A]+S[B]-S[A\cup B]\ra 0$, \ie\ the disconnected surface
$S_{dis}=S[A]+S[B]$ has lower area than the connected surface
$S_{conn}=S[A\cup B]$. Assuming an approximate global pure state,
we are considering the complementary region as the 2-interval
region $(\Sigma_{rad} \cup I)^{c} = [a'_{+},b'_{+}]\cup [b'_{-},a'_{-}]$.
When the future boundary interval $[b'_{+}, b'_{-}]$
is large approaching the entire future boundary, the two intervals
are well-separated (as seen from Figure~\ref{figfuture boundary}), so
the cross-ratio above is indeed small, justifying the use of
(\ref{multiple interval}) for our purposes\
(we have $1-x = \frac{d[a'_{+},a'_{-}]\, d[b_{+}',b_{-}']}
{d[a'_{+},b_{-}']\,d[a'_{-},b_{+}']}\sim 1$ here).
In this limit of approaching the entire future boundary, we have
large $b_\pm'$, amounting to the assumption
(\ref{b'-rD>>rs-a', b>>rs}) here. Note that there is no holography
here:  we are simply applying the island rule in the 2-dim background
obtained from reduction of the $SdS_4$ geometry and looking for
self-consistent island configurations, assuming an
approximate global pure state in the very low de Sitter temperature
limit. It is also worth noting that while the complementary
2-interval region is unambiguously defined, the 3-interval region
is more ambiguous. For instance, in Figure~\ref{figfuture boundary},
one might imagine defining a global Cauchy slice as the spacelike
slice passing through the points
$\big\{{(a_-'+a_+')_{_L}\over 2},\, a_+',\, b_+',\, b_-',\, a_-',\,
{(a_-'+a_+')_{_R}\over 2}\big\}$, where
the left/right endpoints are the approximate midpoints of the
left/right islands (and this ``unit cell'' repeats indefinitely along
the Penrose diagram). Then the 2-interval subregion\ 
$(\Sigma_{rad} \cup I)^{c} = [a'_{+},b'_{+}]\cup [b'_{-},a'_{-}]$\ is
complementary to the 3-interval subregion\
$\Sigma_{rad} \cup I = [{(a_-'+a_+')_{_L}\over 2},\, a_+']\cup
[b'_{+},b'_{-}]\cup [a_-',\, {(a_-'+a_+')_{_R}\over 2}]$ on this
Cauchy slice. It would appear that there is nothing sacrosanct in
choosing these midpoints ${(a_-'+a_+')_{_{L,R}}\over 2}$
to define the slice, whereas the 2-interval complement is well-defined
via the radiation region and island endpoints.
It would be interesting to understand this more elaborately.


In light of the above, the entanglement entropy for the complementary
2-interval region $[a'_{+},b'_{+}]\cup [b'_{-},a'_{-}]$ using
(\ref{multiple interval}) is
\be\label{S_matter 2}
S_{matter}=\frac{c}{3}\,\log\, \frac{d[a'_{+},a'_{-}]\, d[b_{+}',b_{-}']\,
  d[a'_{+},b_{+}']\, d[a'_{-},b_{-}']}{d[a'_{+},b_{-}']\,d[a'_{-},b_{+}']}\ .
\ee
In detail, 
using the Kruskal coordinates (\ref{SdS-XT in FP}), (\ref{SdS-XT in BHP}),
the total generalized entropy (\ref{S_matter 2}) becomes
\begin{align}\label{Stotal=area+log.a'+a'-b'+b'-.a'+b'+a'-b'-/a'+b'-a'-b'+}
    S_{total} & =\frac{2\pi {a'}^{2}}{G_{_N}}+\frac{c}{6} \log \Big[\frac{2^4\lambda^{2}}{\alpha_{D}^2\alpha_{S}^2}(r_{S}-a')(b'-r_{S})(\frac{r_{D}-a'}{l})(\frac{b'-r_{D}}{l})
    \cdot \nn\\
    & \qquad\qquad\qquad\qquad
    (\frac{a'+r_{S}+r_{D}}{l})(\frac{b'+r_{S}+r_{D}}{l}) \sinh^2{X_{a'}} \sinh^2{X_{b'}}\Big]\nn\\
    & \qquad\qquad\qquad + \frac{c}{3} \log \Big[1-2\frac{(r_{S}-a')^{\alpha_{S}\beta_{S}}}{(b'-r_{D})^{\alpha_{D}\beta_{D}}}\,C(a')\, \cosh{(X_{a'}+X_{b'})}\Big]\nn\\
    & \qquad\qquad\qquad -\frac{c}{3} \log \Big[1-2\frac{(r_{S}-a')^{\alpha_{S}\beta_{S}}}{(b'-r_{D})^{\alpha_{D}\beta_{D}}}\,C(a')\, \cosh{(X_{a'}-X_{b'})}\Big]\ ,
\end{align}
where we have added the area term, and $C(a')$ is defined as
\be\label{C(a')rDab}
C(a') = \frac{(b'-r_{s})^{\alpha_{D}\beta_{S}}(a'+r_{S}+r_{D})^{\alpha_{S}\beta_{M}}(b'+r_{S}+r_{D})^{\alpha_{D}\beta_{M}}}{(r_{D}-a')^{\alpha_{S}\beta_{D}}}\ .
\ee
(Note from (\ref{beta}) that $C(a')$ is dimensionless.)
See Appendix \ref{App:Island} for details of this calculation. 

Extremizing (\ref{Stotal=area+log.a'+a'-b'+b'-.a'+b'+a'-b'-/a'+b'-a'-b'+})
with respect to the location of the island boundary $a'$ as
$\frac{\partial S_{total}}{\partial a'}=0$ gives
\begin{align}\label{extremize a'}
&\frac{4 \pi a'}{G_{_N}}+ \frac{c}{6}\Big[-\frac{1}{r_S-a'}-\frac{1}{r_D-a'}+\frac{1}{a'+r_S+r_D}\Big] \nn\\ 
&\quad -\frac{c}{3}{\frac{C(a')}{\sqrt{b'-r_D}\sqrt{r_S-a'}}}\,  \Big[-1+2(r_S-a')(\frac{\alpha_{S}\beta_{M}}{a'+r_{S}+r_{D}}+\frac{\alpha_{S}\beta_{D}}{r_D-a'})\Big]\cdot \nn\\ 
&\qquad\qquad\qquad\quad
\Big[\frac{\cosh({X_{a'}+X_{b'}})}{1-2\frac{(r_{S}-a')^{\alpha_{S}\beta_{S}}}{(b'-r_{D})^{\alpha_{D}\beta_{D}}}\,C(a')\, \cosh{(X_{a'}+X_{b'})}} \nn\\ 
&\qquad\qquad\qquad\qquad
- \frac{\cosh({X_{a'}-X_{b'}})}{1-2\frac{(r_{S}-a')^{\alpha_{S}\beta_{S}}}{(b'-r_{D})^{\alpha_{D}\beta_{D}}}\,C(a')\, \cosh{(X_{a'}-X_{b'})}}\Big] =0\ .
\end{align}

Here, since $r_{D}$ is large, the terms scaling
as $O({1\over r_D})$ can be ignored: thus the
${1\over a'+r_S+r_D}$ and ${1\over r_D-a'}$ are suppressed relative to
${1\over r_S-a'}$\,. With these approximations, (\ref{extremize a'}) becomes
\begin{align}\label{simplified extremize}
&\frac{4 \pi a'}{G_{_N}}- \frac{c}{6}\frac{1}{r_S-a'}+ \frac{c}{3}{\frac{C(a')}{\sqrt{b'-r_D}\sqrt{r_S-a'}}}\, \cdot\Big[\frac{\cosh({X_{a'}+X_{b'}})}{1-2\frac{(r_{S}-a')^{\alpha_{S}\beta_{S}}}{(b'-r_{D})^{\alpha_{D}\beta_{D}}}\,C(a')\, \cosh{(X_{a'}+X_{b'})}} \nn\\ 
&\qquad\qquad\qquad\qquad\qquad\qquad - \frac{\cosh({X_{a'}-X_{b'}})}{1-2\frac{(r_{S}-a')^{\alpha_{S}\beta_{S}}}{(b'-r_{D})^{\alpha_{D}\beta_{D}}}\,C(a')\, \cosh{(X_{a'}-X_{b'})}}\Big] =0\ .
\end{align} 
Next, extremizing (\ref{Stotal=area+log.a'+a'-b'+b'-.a'+b'+a'-b'-/a'+b'-a'-b'+}) with respect to $X_{a'}$ as $\frac{\partial S_{total}}{\partial X_{a'}}=0$ gives
\begin{align}\label{extremize X_a'}
&\coth{X_{a'}}= 2 \sqrt{\frac{r_S-a'}{b'-r_D}} C(a') \cdot \Big[\frac{\sinh({X_{a'}+X_{b'}})}{1-2\frac{(r_{S}-a')^{\alpha_{S}\beta_{S}}}{(b'-r_{D})^{\alpha_{D}\beta_{D}}}\,C(a')\, \cosh{(X_{a'}+X_{b'})}} \nn\\
&\qquad\qquad\qquad\qquad\qquad\qquad\qquad
- \frac{\sinh({X_{a'}-X_{b'}})}{1-2\frac{(r_{S}-a')^{\alpha_{S}\beta_{S}}}{(b'-r_{D})^{\alpha_{D}\beta_{D}}}\,C(a')\, \cosh{(X_{a'}-X_{b'})}}\Big]
\end{align}
We will consider all possible
conditions between $X_{a'}$ and $X_{b'}$ in the extremization
equations to look for consistent solutions to the
location of island boundary, \ie\ the value of $a'$ and $X_{a'}$.

We will first consider, $X_{a'}=X_{b'} \Rightarrow X_{a'}-X_{b'}=0$ and
$X_{a'}+X_{b'}=2X_{a'}$ for large $X_{a'}$ and $X_{b'}$. Then
(\ref{extremize X_a'}) becomes 
\begin{equation}\label{final X_a' eq}
1-2 \sqrt{\frac{r_S-a'}{b'-r_{D}}} C(a')\cosh{2X_{a'}}=2\sqrt{\frac{r_S-a'}{b'-r_{D}}} C(a') \frac{\sinh{2X_{a'}}}{\coth{X_{a'}}}
\end{equation}
Putting this condition (\ref{final X_a' eq}) back in
(\ref{simplified extremize}) gives
\begin{align}\label{final a' eq}
&\frac{4 \pi a'}{G_{_N}}- \frac{c}{6}\frac{1}{r_S-a'}+ \frac{c}{3}{\frac{C(a')}{\sqrt{b'-r_D}\sqrt{r_S-a'}}}\,\Big[\frac{\cosh(2{X_{a'}})}{2\frac{(r_{S}-a')^{\alpha_{S}\beta_{S}}}{(b'-r_{D})^{\alpha_{D}\beta_{D}}}\,C(a')\, \frac{\sinh{2X_{a'}}}{\coth{X_{a'}}}}-1\Big]=0 \nn\\
\Rightarrow\quad &\frac{4 \pi a'}{G_{_N}}- \frac{c}{3}{\frac{C(a')}{\sqrt{b'-r_D}\sqrt{r_S-a'}}}- \frac{c}{6} \frac{1}{r_{S}-a'}\Big(1-\frac{\coth{X_{a'}}}{\tanh{2X_{a'}}}\Big)=0\,.
\end{align}
For large $X_{a'}$ and $X_{b'}$, the third term in (\ref{final a' eq}) is
small compared to the second term. So we can ignore the third term and
(\ref{final a' eq}) becomes
\be\label{a' eq}
a' \,\simeq\, {1\over \sqrt{r_{S}-a'}} \ \frac{G_{_N}c}{12\pi}\,
\frac{1}{\sqrt{b'-r_{D}}}\, C(a')\, . 
\ee
Now we recall that we are in the semiclassical regime where
\be
0 \ll c \ll {1\over G_{_N}}\ ,
\ee
so that the classical area term in the generalized entropy is dominant
but the bulk matter makes nontrivial subleading contributions (which are
not so large as to cause significant backreaction on the classical
geometry).

We are looking for an island with boundary $a'\sim r_S$ near the black
hole horizon: this corroborates with the fact that since the entire
right hand side in (\ref{a' eq}) is $O(G_{_N}c)$, in the classical
limit $G_{_N}c=0$ we
obtain\ $a'\sqrt{r_s-a'}\simeq O(G_{_N}c)\sim 0$ giving $a'=r_S$, \ie\
the quantum extremal surface localizes on the black hole horizon. Thus
we can solve the above extremization equation in perturbation theory
setting $a'\sim r_S$ at leading order to find the first order
correction in $G_{_N}c\ll 1$\,: then schematically we have
\be\label{a' final eq}
r_{S}-a' \simeq \frac{K^{2}}{r_{S}^{2}} \frac{1}{b'-r_{D}}\ ,\qquad
K=\frac{G_{_N}c}{12\pi}\,C(r_S)\ ,
\ee
Thus, we finally obtain (with $C(r_S)$ from (\ref{C(a')rDab})
setting $a'=r_S$)
\be\label{a' value}
a'\simeq r_{S}- \frac{(G_{_N}c)^{2}}{144 \pi^{2} r_{S}^2(b'-r_{D})}\,
C(r_S)^2.
\ee
Solving now for $X_{a'}$ from (\ref{extremize X_a'}), we obtain
\be\label{final X_a' value}
\cosh{2X_{a'}}=\frac{6\pi}{G_N c} \frac{r_S (b'-r_D)}{C(r_S)^{2}}\frac{\coth{X_{a'}}}{\tanh{2X_{a'}}+\coth{X_{a'}}}\ .
\ee
This is a large $X_{a'}$ value with $e^{2X_{a'}}$ scaling approximately as
$O({1\over G_{_N}c})$. Considering $X_{a'}\sim -X_{b'}$ does not yield
consistent island solutions: see App.~\ref{wrong2}. Further,
considering potential island solutions just outside the horizon
turns out to be inconsistent: see App.~\ref{wrong1}.
Thus (\ref{a' value}), (\ref{final X_a' value}), with $X_{a'}\sim X_{b'}$,
encode the correct island solution for the future boundary radiation
region. The condition $X_{a'}=X_{b'}$ is consistent with the expectation
that the island location lies on the same Cauchy slice as the radiation
region location (along the same lines as the condition $t_a=t_b$ within
the static diamond in \cite{Goswami:2022ylc})). This amounts to the
requirement of spacelike separation in considering the island and
radiation as an effectively single entity which purifies so the fact
that we recover this is not surprising. The fact that islands outside
the horizon are inconsistent is due to causality: the entanglement
wedge cannot lie within the causal wedge (we explain this further in
the Discussion sec.~\ref{conclusions}).

With the value of $a'$ in (\ref{a' value}) and $X_{a'}$ in
(\ref{final X_a' value}), the total on-shell entanglement entropy in
(\ref{Stotal=area+log.a'+a'-b'+b'-.a'+b'+a'-b'-/a'+b'-a'-b'+}) becomes
\begin{align}\label{S on shell}
S_{o.s}=\  &  \frac{2 \pi r_{S}^2}{G_{_N}}
  - \frac{c^{2}G_{_N}}{36\pi r_{S}(b'-r_{D})}\,C(r_S)^2 \\
+\, & \frac{c}{6} \log\Big[\frac{16\lambda^2\beta_{s}^2\beta_{D}^2(b'-r_{D})^{2}}{l^{4}}\,   { (r_{D}-r_{S})^{1+2\alpha_{S}\beta_{D}}\over {(b'-r_S)^{2\alpha_{S}\beta_{D}-1}(2r_{S}+r_{D})^{2\alpha_{S}\beta_{M}-1}(b'+r_{S}+r_{D})^{2\alpha_{S}\beta_{M}-1}}}\, 
\Big]\,. \nn 
\end{align}
which is independent of length $X_{a'}$ and $X_{b'}$, stemming from the
presence of the island.
The leading first (area) term is twice the Bekenstein-Hawking entropy
of the black hole, while the subleading second and third terms arising
from the bulk entropy of the radiation region purified by the island
are constant terms not growing in length.  This recovers the
expectations on the Page curve for the entropy of the bulk matter or
Hawking radiation considered near the future boundary. The bulk
matter at the future boundary radiation region is entangled with the
island-like region located just inside the black hole interior in
these semiclassical approximations at very low ambient de Sitter
temperature.

Comparing the entanglement entropy without the island
(\ref{approx.entropy}) and that with the island (\ref{S on shell})
provides the critical length $X_{Page}$ at which the island transition
occurs: we obtain
\be\label{Page length}
\frac{c}{3} X_{Page} \,\sim \, 2 S_{BH}
\qquad\Rightarrow\qquad
X_{Page} \, \sim \, \frac{6\pi {r_{S}}^2}{G_{N}c}\,.
\ee
The entropy with the island alongwith the associated purification
is lower and dominates over the no-island configuration beyond this 
critical length $X_{Page}$. Note that here the critical Page length
$X_{Page}$ is a dimensionless quantity, using (\ref{AC of b}). This
then corresponds to a Page time $t_P\sim \beta_D X_{Page}$ which
using (\ref{beta}) and the approximations (\ref{m<<l}) gives
the large value $t_P\sim lS_{BH}$\ (note that this uses the cosmological
Kruskal coordinates, distinct from the black hole Kruskal coordinates
in \cite{Goswami:2022ylc}).  It is however important to note that
this Page length is much smaller than another potentially relevant
quantity $X_{_P}^{dS}\sim S_{dS}$ controlled by the entropy
$S_{dS}$ of the cosmological horizon.
In the small black hole limit (\ref{m<<l}) we are considering,
$S_{dS}\sim {l^2\over G_{_N}}\ \gg\ S_{BH}\sim {r_S^2\over G_{_N}}$\,,
and we do not see any effects above, stemming from the ambient de Sitter
space which is just a frozen background. So our critical Page length
(\ref{Page length}) controlled by black hole entropy alone is
consistent with the separation of scales in the limit (\ref{m<<l}).
Away from this limit, the black hole horizon shrinks while the
cosmological horizon absorbs and grows, resulting in a nontrivial
nonequilibrium system. It would of course be interesting to
understand de Sitter horizon physics, but this appears substantially
more challenging within our framework.

Finally, it is worth noting that there are also timelike separated
quantum extremal surface solutions following from the extremization of
the generalized entropy with respect to the future boundary observer:
we discuss these solutions in App.~\ref{timelike separated qes}.  The
timelike separation implies that the on-shell generalized entropy
becomes complex valued. While complex entropies are known in
investigations in pure de Sitter space (which does not have a
sufficiently wide Penrose diagram) and suggest new objects
\cite{Narayan:2015vda}, \cite{Sato:2015tta}, \cite{Narayan:2017xca},
\cite{Doi:2022iyj}, \cite{Narayan:2022afv}, it is consistent to ignore
them in the Schwarzschild de Sitter context where spacelike separated
quantum extremal surfaces do exist in accord with physical Page curve
expectations for the black hole information paradox.

\section{Discussion}\label{conclusions}

We have studied small 4-dim Schwarzschild de Sitter black
holes in the limit of very low de Sitter temperature, building further
on previous work \cite{Goswami:2022ylc} for observers within the
static diamond far from the black hole horizon. In the present work,
we have been considering the black hole Hawking radiation in a
radiation region interval at the future boundary (see
Figure~\ref{figfuture boundary}).  The black hole mass is adequately
large that quasistatic approximations to the evaporating black hole in
semiclassical gravity are valid. We assume the black hole radiation
approximated as a 2-dim CFT at nearly zero temperature propagating in
a 2-dim dilaton gravity background (\ref{SdS-2d}) obtained by
dimensional reduction of the 4-dim spacetime. Including appropriate
island contributions, we find that the generalized entropy satisfies
expectations from the Page curve for the evolution of bulk matter near
the future boundary. Our analysis has parallels with \cite{Svesko:2022txo}
which studied island resolutions for $dS_{2}$ JT gravity with regard
to the future boundary. Our setup here is somewhat more complicated since the
assumption of an approximate global pure state is only reasonable, if
at all, at very low de Sitter temperature. The fact that these
approximate calculations vindicate the island paradigm perhaps
suggests the existence of better, more fundamental ways to formulate
the information paradox in such nontrivial gravitational backgrounds
and of deeper insights into replica wormholes in these sorts of
quasistatic gravitational backgrounds.

The Schwarzschild de Sitter ($SdS$) black hole is unstable and thus
somewhat different from the $AdS$ black hole. In our small black hole
limit (\ref{m<<l}) the ambient de Sitter space effectively remains a
frozen background reservoir. In a quasistatic approximation, the black
hole evaporates away slowly, and our analysis using the eternal $SdS$
black hole shows the radiation entanglement entropy including the
island becoming saturated at some finite value (\ref{S on shell}),
approximately $2S_{BH}$ (so the Page curve saturates rather than
falls). As the black hole evaporates, its entropy decreases so the
saturation value of the radiation entropy decreases leading to the
black hole Page curve falling slowly in accord with the approximately
pure state that the black hole formed from. Strictly speaking, the
ambient de Sitter space temperature (albeit much lower than that of
the black hole) implies that the pure state consideration is just an
approximation.  It would be interesting to study the $SdS$ black hole
modelling the bulk matter CFT in the thermal state at finite de Sitter
temperature.

We recall that in \cite{Goswami:2022ylc}, the radiation region was
within the static diamond (with endpoints $b_+$ or $b_-$ in the left
or right static diamond, in Figure~\ref{figfuture boundary}). The late
time island location was then found to be within the static diamond,
just outside the black hole horizon in that case. As we have seen, the
island location we have found currently is just inside the black hole
horizon, which at first sight might seem contradictory. However this
is in fact consistent in the current case. First, the future boundary
interval $(b_+', b_-')$ in the present case receives Hawking radiation
from both the left and right black hole patches, propagating past the
left and right cosmological horizons bounding the left and right
static diamonds.  So this setup is physically distinct from the
previous case of a single static diamond. Secondly, in obtaining the
island locations we have been considering the limit of large $X_{a'},
X_{b'}$, in the extremization equations. In this limit the future
boundary interval $(b_+', b_-')$ approaches the entire future
boundary, \ie\ the points $b_\pm'$ approach the endpoints of
$I^+$. Note that the left and right static diamonds are now within the
causal wedge of this interval. It would then be causally inconsistent
for the entanglement wedge to be within the causal wedge of the
radiation interval. The entanglement wedge of the radiation region is
the domain of dependence, or bulk causal diamond of the spacelike
surface between the boundary of the radiation region and the island
boundary (location of quantum extremal surface). As it stands, the
island boundary is just inside the black hole horizon so it lies
outside the causal wedge, nicely avoiding inconsistency.

The black hole interior island solution is ultimately supported by the
calculational fact that other possibilities lead to inconsistencies:
for instance, blindly looking for island solutions outside the horizon
in the present case exhibits inconsistency in the extremization
equations. We carried out this exercise by performing the calculation
of sec.~\ref{sec:EEisland} using static diamond Kruskal coordinates
for the potential island lying just outside the black hole horizon in
the static diamond ($a'\gtrsim r_s$ in this case, somewhat akin to the
parametrizations in \cite{Goswami:2022ylc} reviewed in
App.~\ref{detailed calculation of previous paper}). The analog of
(\ref{Stotal=area+log.a'+a'-b'+b'-.a'+b'+a'-b'-/a'+b'-a'-b'+}) in this
case leads to extremization equations similar to (\ref{extremize a'})
and (\ref{extremize X_a'}): however there are subtle differences which
ensure that the analogs of (\ref{final X_a' eq}) and (\ref{final a'
  eq}) together do not give consistent island solutions (see
App.~\ref{wrong1}). Further, as we also noted after (\ref{final X_a'
  value}), potential island solutions with $X_{a'}=-X_{b'}$ (instead
of $X_{a'}=X_{b'}$) also lead to inconsistencies (App.~\ref{wrong2}).
Thus overall, our semiclassical island solution in (\ref{a' value})
and (\ref{final X_a' value}) should be regarded as nontrivial. Perhaps
the self-consistency of these calculations (in particular using the
complementary 2-interval bulk matter entropy) also vindicates the
assumption of approximate purity of the initial matter that made the
black hole in this very low temperature de Sitter ambience. It would
be interesting to explore this in more detail, as discussed around
(\ref{multiple interval}).

The separation of scales in the small black hole limit (\ref{m<<l})
ensures that black holes can be regarded as localized subsystems
analyzable by distinct classes of observers (or metaobservers).  Then,
abstracting away from our technical analysis in Schwarzschild de
Sitter black holes vindicates some general lessons for the black hole
information paradox here as well. Islands appear to emerge
self-consistently evading paradoxes with (i) unitarity as encapsulated
by the Page curve (late static patch times and large future boundary
lengths), (ii) causality (the island boundary does not lie within the
causal wedge), (iii) overcounting (the purifying island is spacelike
separated from the radiation, lying on the same Cauchy slice).
In this light, de Sitter space itself and cosmological horizons appear
exotic: extremal surfaces anchored at the future boundary 
involve timelike separations (\eg\ \cite{Chen:2020tes},
\cite{Goswami:2021ksw}, for quantum extremal surfaces, and
\cite{Narayan:2015vda}, \cite{Sato:2015tta}, \cite{Narayan:2017xca},
\cite{Doi:2022iyj}, \cite{Narayan:2022afv} for classical RT/HRT
surfaces). So de Sitter space, and perhaps cosmology more generally,
require new insights.

Our discussions of Schwarzschild de Sitter are entirely within the
bulk framework of semiclassical gravity, with no holography per se
(except in the broad sense of gravity being intrinsically
holographic).  The future boundary is well-defined as a place where
gravity is manifestly weak: however we have simply applied the island
formulation in these relatively complicated higher dimensional models
under various assumptions and approximations without rigorous
justification.  So this appears to stretch the regimes of validity of
the original island proposals, although it corroborates the general
expectations laid out in \cite{Almheiri:2020cfm}.  It would be nice to
better understand in more fundamental ways the deeper underpinnings of
semiclassical gravity that encode these self-consistent island
formulations of the black hole information paradox. In this regard
it might be interesting to understand the interplay between the generalized
entropy and its extremization and gravity actions (see \eg\
\cite{Pedraza:2021cvx}, \cite{Pedraza:2021ssc}, \cite{Morvan:2022ybp},
\cite{Svesko:2022txo}) in the context of the general 2-dim dilaton gravity
theories (\ref{SdS-2d}) we consider here arising from reduction of $SdS_4$.

\vspace{6mm}

{\footnotesize {\bf Acknowledgements:}\ \ \ It is a pleasure
  to thank Ahmed Almheiri, Sitender Kashyap, Alok Laddha, Juan Maldacena,
  Suvrat Raju and Amitabh Virmani for discussions and comments on a draft.
  This work is partially supported by a grant to CMI from the Infosys
  Foundation.}

\vspace{2mm}

\appendix

\renewcommand{\theequation}{\thesection.\arabic{equation}}

\section{Review: static patch radiation entropy with islands }\label{detailed calculation of previous paper}

In this section, we briefly review the calculation in
\cite{Goswami:2022ylc} of the island resolution of the black hole
information paradox in Schwarzschild de Sitter black holes in the
limit of small black hole mass and very low de Sitter temperature.
This has close parallels with islands in flat space Schwarzschild
black holes \cite{Hashimoto:2020cas}.
Considering the radiation region in the static patch, far from the
black hole horizon but within the cosmological horizon, the
entanglement entropy of the bulk matter can be shown to increase
unboundedly. Including an island region $I\equiv [a_-,a_+]$ straddling
across the black hole horizon, we consider the entanglement entropy of
the interval $R_-\cup I\cup R_+$. Strictly speaking, the bulk matter
should be approximated as a CFT at finite temperature corresponding to
the de Sitter temperature: however in the limit of very low de Sitter
temperature and a small mass black hole, we can approximate the bulk
theory to be in an approximately pure state. Then calculating
the complementary interval entropy and appending the area term (from
the island boundary area) gives the total entanglement entropy
\cite{Goswami:2022ylc}
\begin{align}\label{Stotal=area+log.a+a-b+b-.a+b+a-b-/a+b-a-b+}
    S_{total} & =\frac{2\pi a^{2}}{G_{_N}}+\frac{c}{6} \log \Big[\frac{2^8r_{S}^4}{(\frac{r_{D}-r_{S}}{l})^4(\frac{2r_{S}+r_{D}}{l})^4}(a-r_{S})(b-r_{S})(\frac{a-r_{D}}{l})(\frac{b-r_{D}}{l})
    \cdot \nn\\
    & \qquad\qquad\qquad\qquad
    (\frac{a+r_{S}+r_{D}}{l})(\frac{b+r_{S}+r_{D}}{l}) \cosh^2{\frac{t_{a}}{2\beta_{S}}} \cosh^2{\frac{t_{b}}{2\beta_{S}}}\Big]\nn\\
    & \qquad\qquad\qquad + \frac{c}{3} \log \Big[1-2\frac{(a-r_{S})^{\alpha_{S}\beta_{S}}}{(b-r_{S})^{\alpha_{S}\beta_{S}}}\,C(a)\, \cosh{\big(\alpha_{S}(t_{a}-t_{b})\big)}\Big]\nn\\
    & \qquad\qquad\qquad -\frac{c}{3} \log \Big[1+2\frac{(a-r_{S})^{\alpha_{S}\beta_{S}}}{(b-r_{S})^{\alpha_{S}\beta_{S}}}\,C(a)\, \cosh{\big(\alpha_{S}(t_{a}+t_{b})\big)}\Big]\ ,
\end{align}
where $C(a)$ is 
\be\label{C(a)rDab}
C(a) = \frac{(r_{D}-b)^{\alpha_{S}\beta_{D}}(a+r_{S}+r_{D})^{\alpha_{S}\beta_{M}}}{(r_{D}-a)^{\alpha_{S}\beta_{D}}(b+r_{S}+r_{D})^{\alpha_{S}\beta_{M}}}\ .
\ee
Extremizing $S_{total}$ in (\ref{Stotal=area+log.a+a-b+b-.a+b+a-b-/a+b-a-b+})
with respect to the location of the island boundary $a$ gives
\begin{align}\label{simplified extremize previous}
&\frac{4 \pi a}{G_{_N}}+ \frac{c}{6}\frac{1}{a-r_S}- \frac{c}{3}{\frac{C(a)}{\sqrt{b-r_S}\sqrt{a-r_S}}}\, \,\Big[\frac{\cosh(\alpha_S(t_{a}-t_{b}))}{1-2\frac{(a-r_{S})^{\alpha_{S}\beta_{S}}}{(b-r_{S})^{\alpha_{S}\beta_{S}}}\,C(a)\, \cosh{(\alpha_S(t_{a}-t_{b}))}} \nn\\
&\qquad\qquad\qquad\qquad\qquad\qquad + \frac{\cosh{(\alpha_S(t_{a}+t_{b})))}}{1+2\frac{(a-r_{S})^{\alpha_{S}\beta_{S}}}{(b-r_{S})^{\alpha_{S}\beta_{S}}}\,C(a)\, \cosh{(\alpha_S(t_{a}+t_{b}))}}\Big] =0\ .
\end{align} 
Next, extremizing $S_{total}$ from
(\ref{Stotal=area+log.a+a-b+b-.a+b+a-b-/a+b-a-b+}) with respect to $t_{a}$
gives
\begin{align}\label{extremize t_a}
&\tanh{(\alpha_{S}t_a)}= 2 \sqrt{\frac{a-r_S}{b-r_S}} C(a) \alpha_S \, \Big[\frac{\sinh(\alpha_S(t_{a}-t_{b}))}{1-2\frac{(a-r_{S})^{\alpha_{S}\beta_{S}}}{(b-r_{S})^{\alpha_{S}\beta_{S}}}\,C(a)\, \cosh{(\alpha_S(t_{a}-t_{b}))}} \nn\\
&\qquad\qquad\qquad\qquad\qquad\qquad + \frac{\sinh(\alpha_S(t_{a}+t_{b}))}{1+2\frac{(a-r_{S})^{\alpha_{S}\beta_{S}}}{(b-r_{S})^{\alpha_{S}\beta_{S}}}\,C(a)\, \cosh{(\alpha_S(t_{a}+t_{b}))}}\Big]
\end{align}
First consider $t_{a}=t_{b}$ so $t_{a}-t_{b}=0$ and $t_{a}+t_{b}=2t_{a}$
for large $t_{a}, t_{b}$. Then (\ref{extremize t_a}) becomes
\begin{equation}\label{final t_a eq}
1+2 \sqrt{\frac{a-r_S}{b-r_{S}}} C(a)\cosh{(\alpha_S \cdot 2t_b)}=2\sqrt{\frac{a-r_S}{b-r_{S}}} C(a) \frac{\sinh{(\alpha_S \cdot 2t_b)}}{\tanh{(\alpha_S t_b)}}
\end{equation}
Next, putting this condition (\ref{final t_a eq}) back in
(\ref{simplified extremize previous}) gives
\begin{align}\label{final a eq}
&\frac{4 \pi a}{G_{_N}}+ \frac{c}{6}\frac{1}{a-r_S}- \frac{c}{3}{\frac{C(a)}{\sqrt{b-r_S}\sqrt{a-r_S}}}\,\Big[1+\frac{\cosh{(\alpha_S\cdot 2t_b)}}{2\frac{(a-r_{S})^{\alpha_{S}\beta_{S}}}{(b-r_{S})^{\alpha_{S}\beta_{S}}}\,C(a)\, \frac{\sinh{(\alpha_S\cdot 2t_b)}}{\tanh{(\alpha_S t_b)}}}\Big]=0 \nn\\
\Rightarrow\ \ \
&\frac{4 \pi a}{G_{_N}}- \frac{c}{3}{\frac{C(a)}{\sqrt{b-r_S}\sqrt{a-r_S}}}+ \frac{c}{6} \frac{1}{a-r_{S}}\Big(1-\frac{\tanh{(\alpha_S t_b)}}{\tanh{(\alpha_S\cdot 2t_b)}}\Big)=0\,.
\end{align}
For large $t_{a}$ and $t_{b}$, the third term in (\ref{final a eq}) is
small relative to the second term. So we can ignore the third term and
(\ref{final a eq}) becomes
\be\label{a eq}
a \,\simeq\, {1\over \sqrt{a-r_{S}}} \ \frac{G_{_N}c}{12\pi}\,
\frac{1}{\sqrt{b-r_{S}}}\, C(a)\, . 
\ee
Thus solving this in perturbation theory for the first order
correction in $G_{_N}c\ll 1$ gives
\be\label{a value}
a\simeq r_{S}+ \frac{(G_{_N}c)^{2}}{144 \pi^{2} r_{S}^2(b-r_{S})}\,
C(r_S)^2\,,
\ee
setting $a\sim r_S$ in $C(a)$ etc. Solving for $t_a$ from
(\ref{extremize t_a}), we obtain
\be\label{final t_a value}
\cosh{(\alpha_S \cdot 2t_a)}=\frac{6\pi}{G_N c} \frac{r_S (b-r_S)}{C(r_S)^{2}}\frac{\tanh{(\alpha_S t_a)}}{\tanh{(\alpha_S \cdot 2t_a)}-\tanh{(\alpha_S t_b)}}
\ee
Equations (\ref{a value}), (\ref{final t_a value}), at late times
$t_a=t_b$, recover the result in \cite{Goswami:2022ylc}.\
(Considering $t_{a}=-t_{b}$ \ie\ $t_{a}+t_{b}=0$ and $t_{a}-t_{b}=2t_{a}$
for large $t_{a}$ and $t_{b}$, the above analysis can be seen to give
physically inconsistent solutions.)\
The island is a little outside the horizon. The late time generalized
entropy including the island contribution is finite, approximately
twice the black hole entropy upto small corrections from the bulk
matter.

\section{Details: entropy in the no-island case}\label{App:noIsland}

This section contains some details on the calculations of entanglement
entropy in the absence of the island in sec.~\ref{sec:noIsl}.\
Using (\ref{SdS-XT in FP}) and calculating each part of $S_{matter}$ in
(\ref{matter entropy}) separately gives
\be\label{A.38}
U_{D_-}'-U_{D_+}'= -e^{-T_{b'}}\cdot 2\sinh{X_{b'}}\,,\qquad
V_{D_+}'-V_{D_-}'= -e^{-T_{b'}}\cdot 2\sinh{X_{b'}}\,,
\ee
\be\label{A.40}
W_{b'_{+}}=W_{b'_{-}}= \sqrt{b'}\,l\,\alpha_{D}\,(b'-r_{D})^{-\frac{1-2\alpha_{D}\beta_{D}}{2}}(b'-r_{S})^{-\frac{1+2\alpha_{D}\beta_{S}}{2}} (b'+r_{S}+r_{D})^{-\frac{1+2\alpha_{D}\beta_{M}}{2}}\ ,
\ee
\begin{align}\label{A.41}
W_{b'_{+}}W_{b'_{-}} & =b'\,l^{2}\,\alpha_{D}^2 (b'-r_{D})^{-(1-2\alpha_{D}\beta_{D})} (b'+r_{S}+r_{D})^{-(1+2\alpha_{D}\beta_{M})} (b'-r_{S})^{-(1+2\alpha_{D}\beta_{S})}\nn\\
&= \frac{\alpha_{D}^2}{\lambda b' |f(b')|} e^{-2T_{b'}} \ .
\end{align}
Plugging all these into (\ref{matter entropy}) gives\
$S_{matter} = \frac{c}{6} \log[\frac{\lambda b'}{\al_{D}^2} e^{2T_{b'}} |f(b')| e^{-2T_{b'}}   4\sinh^2{X_{b'}}]$,\ \ie\
\begin{align}\label{A.42}
   S_{matter} =  \frac{c}{6} \log\Big[\frac{\lambda}{l^2 \al_{D}^2}(b'-r_{S}) \, (b'-r_D)(b'+r_{S}+r_{D})\, 4\sinh^2{X_{b'}}\Big]\,.
\end{align}
Thus finally, we obtain (\ref{Total entropy}).

\section{Details: late-time entropy with island}\label{App:Island}

Here we give details on sec.~\ref{sec:EEisland}.
We are looking to calculate (\ref{S_matter 2}), \ie\
\be\label{B.43}
    S_{matter}=\frac{c}{3}\log \frac{d(a'_{+},a'_{-})d(b_{+}',b_{-}')d(a'_{+},b_{+}')d(a'_{-},b_{-}')}{d(a'_{+},b_{-}')d(a'_{-},b_{+}')}\ .
\ee
Now calculating each part in $S_{matter}$ separately\\
\be\label{B.44}
    \log[d(a'_{+},a'_{-})]=\frac{1}{2}\log \frac{[(U'_{S_-}-U'_{S_+})(V'_{S_+}-V'_{S_-})]}{W'_{a_{+}'}W'_{a_{-}'}}
\ee
with $W_{a'}$ as in (\ref{SdS-kruskal in BHP}).    
Then we have
\be\label{B.45}
U'_{S_-}-U'_{S_+}= e^{T_{a'}}\cdot 2 \sinh{X_{a'}}\,,\qquad 
V'_{S_+}-V'_{S_-}= e^{T_{a'}}\cdot 2 \sinh{X_{a'}}\,,
\ee
\begin{align}
W_{a'_{+}}W_{a'_{-}} & =a'l^{2}\alpha_{S}^2 (r_{D}-a')^{-(1+2\alpha_{S}\beta_{D})} (a'+r_{S}+r_{D})^{(2\alpha_{S}\beta_{M}-1)} (r_{S}-a')^{(2\alpha_{S}\beta_{S}-1)} \nn \\
& = \frac{\alpha_{S}^2}{\lambda a' |f(a')|} e^{2T_{a'}} \ .
\end{align}
Putting all these expressions together in (\ref{B.44}) gives
\begin{align}\label{B.48}
\log[d(a'_{+},a'_{-})] & = \frac{1}{2} \log\Big[\frac{\lambda a'}{\al_{S}^2} e^{-2T_{a'}} |f(a')| e^{2T_{a'}}   4\sinh^2{X_{a'}}\Big] \nn \\
&=\frac{1}{2}\log \Big[\frac{\lambda}{l^2\alpha_{S}^2}(r_{D}-a')(r_{S}-a')(a'+r_{S}+r_{D})\cdot 4\sinh^2{X_{a'}}]\,.
\end{align}
Similarly we obtain
\begin{align}\label{B.49}
    \log[d(b_{+}',b_{-}')] & =\frac{1}{2}\log \Big[\frac{\lambda}{l^2\alpha_{D}^2}(b'-r_{D})(b'-r_{S})(b'+r_{S}+r_{D}) \cdot 4\sinh^{2}{X_{b'}}]\,.
\end{align}
Now, putting (\ref{B.48}) and (\ref{B.49}) together gives
\begin{align}\label{log.a'+a'-b'+b'-}
   {c\over 3} \log[d(a'_{+},a'_{-})d(b_{+}',b_{-}')] 
    & = \frac{c}{6} \log\Big[\frac{2^4 \lambda^2}{\al_{D}^2\al_{S}^2}(r_{S}-a')(b'-r_{S})(\frac{r_{D}-a'}{l})(\frac{b'-r_{D}}{l})
    \cdot \nn\\
    &  \qquad\qquad
    (\frac{a'+r_{S}+r_{D}}{l})(\frac{b'+r_{S}+r_{D}}{l})\sinh^2{X_{a'}}\sinh^2{X_{b'}}\Big]\ .
\end{align}
We next calculate other relevant contributions using (\ref{SdS-XT in FP})
and (\ref{SdS-XT in BHP}): we have
$$U'_{D_+}-U'_{S_+}= e^{(X_{b'}-T_{b'})}-e^{-(X_{a'}-T_{a'})}, \qquad  
  V'_{S_+}-V'_{D_+}= e^{(X_{a'}+T_{a'})}-e^{-(X_{b'}+T_{b'})} \,,$$
$$U'_{D_-}-U'_{S_-}=e^{-(X_{b'}+T_{b'})}-e^{(X_{a'}+T_{a'})}, \qquad V'_{S_-}-V'_{D_-}=e^{-(X_{a'}-T_{a'})}-e^{(X_{b'}-T_{b'})}\,,$$
so
\begin{align}\label{B.51}
    d(a'_{+},b_{+}') &= \frac{1}{\sqrt{W'_{a'_{+}}W'_{b'_{+}}}}\Big[(U'_{D_+}-U'_{S_+})(V'_{S_+}-V'_{D_+})\Big]^{\frac{1}{2}} \nn \\
    &  =\frac{1}{\sqrt{W'_{a'_{+}}W'_{b'_{+}}}} \Bigg[e^{(T_{a'}-T_{b'})}\cdot \Bigg(2\cosh{(X_{a'}+X_{b'})}-2\cosh{(T_{a'}+T_{b'})}\Bigg)\Bigg]^{\frac{1}{2}}\,,
\end{align}
\begin{align}\label{B.52}
    d(a'_{-},b_{-}') & = \frac{1}{\sqrt{W'_{a'_{-}}W'_{b'_{-}}}}\Big[(U'_{D_-}-U'_{S_-})(V'_{S_-}-V'_{D_-})\Big]^{\frac{1}{2}} \nn \\
    & = \frac{1}{\sqrt{W'_{a'_{-}}W'_{b'_{-}}}} \Bigg[e^{(T_{a'}-T_{b'})}\cdot \Bigg(2\cosh{(X_{a'}+X_{b'})}-2\cosh{(T_{a'}+T_{b'})}\Bigg)\Bigg]^{\frac{1}{2}}\,.
\end{align}
Now, putting (\ref{B.51}) and (\ref{B.52}) together
\begin{align}\label{B.53}
  d(a'_{+},b_{+}')\,d(a'_{-},b_{-}')  &=\frac{1}{\sqrt{W'_{a'_{+}}W'_{b'_{+}}W'_{a'_{-}}W'_{b'_{-}}}}\Bigg[e^{(T_{a'}-T_{b'})}\cdot \Bigg(2\cosh{(X_{a'}+X_{b'})}-2\cosh{(T_{a'}+T_{b'})}\Bigg)\Bigg] .
\end{align}
Similarly, we have
$$U_{D_-}'-U'_{S_+}= e^{-(X_{b'}+T_{b'})}-e^{-(X_{a'}-T_{a'})}, \qquad V'_{S_+}-V'_{D_-}= e^{(X_{a'}+T_{a'})}-e^{(X_{b'}-T_{b'})}\,,$$
$$U_{D_+}'-U'_{S_-}= e^{(X_{b'}-T_{b'})}- e^{(X_{a'}+T_{a'})}, \qquad V'_{S_-}-V'_{D_+}= e^{-(X_{a'}-T_{a'})}- e^{-(X_{b'}+T_{b'})}\,,$$
so
\begin{align}\label{B.54}
    d(a'_{+},b_{-}') & =\frac{1}{\sqrt{W'_{a'_{+}}W'_{b'_{-}}}}\Big[(U_{D_-}'-U'_{S_+})(V'_{S_+}-V'_{D_-})\Big]^{\frac{1}{2}} \nn \\
    & = \frac{1}{\sqrt{W'_{a'_{+}}W'_{b'_{-}}}} \Bigg[e^{(T_{a'}-T_{b'})}\cdot \Bigg(2\cosh{(X_{a'}-X_{b'})}-2\cosh{(T_{a'}+T_{b'})}\Bigg)\Bigg]^{\frac{1}{2}}\,,
\end{align}
\begin{align}\label{B.55}
    d(a'_{-},b_{+}') & ={\frac{1}{\sqrt{W'_{a'_{-}}W'_{b'_{+}}}}}\Big[(U_{D_+}'-U'_{S_-})(V'_{S_-}-V'_{D_+})\Big]^{\frac{1}{2}} \nn \\
    & = \frac{1}{\sqrt{W'_{a'_{-}}W'_{b'_{+}}}} \Bigg[e^{(T_{a'}-T_{b'})}\cdot \Bigg(2\cosh{(X_{a'}-X_{b'})}-2\cosh{(T_{a'}+T_{b'})}\Bigg)\Bigg]^{\frac{1}{2}}\,.
\end{align}
Now, putting (\ref{B.54}) and (\ref{B.55}) together
\begin{align}\label{B.56}
    d(a'_{+},b_{-}')\,d(a'_{-},b_{+}')&=\frac{1}{\sqrt{W'_{a'_{+}}W'_{b'_{-}}W'_{a'_{-}}W'_{b'_{+}}}}\Bigg[e^{(T_{a'}-T_{b'})}\cdot \Bigg(2\cosh{(X_{a'}-X_{b'})}-2\cosh{(T_{a'}+T_{b'})}\Bigg)\Bigg] .
\end{align}

Putting (\ref{B.53}) and (\ref{B.56}) together we get
\begin{align}\label{B.57}
    \frac{c}{3} \log \frac{d(a'_{+},b_{+}')d(a'_{-},b_{-}')}{d(a'_{+},b_{-}')d(a'_{-},b_{+}')} & =\frac{c}{3} \log \Big[\frac{2\cosh{(T_{a'}+T_{b'})}-2\cosh{(X_{a'}+X_{b'})}}{2\cosh{(T_{a'}+T_{b'})}-2\cosh{(X_{a'}-X_{b'})}}\Big]\,.
\end{align}
Here
\begin{align}\label{B.58}
     2\cosh{(T_{a'}+T_{b'})}& =\,\frac{1}{2}\,\Big[\frac{(r_{S}-a')^{\alpha_{S}\beta_{S}}(a'+r_{D}+r_{S})^{\alpha_{S}\beta_{M}}(b'-r_{S})^{\alpha_{D}\beta_{S}}(b'+r_{D}+r_{S})^{\alpha_{D}\beta_{M}}}{(r_{D}-a')^{\alpha_{S}\beta_{D}}(b'-r_{D})^{\alpha_{D}\beta_{D}}}
    \nn\\
    & \qquad + \frac{(r_{D}-a')^{\alpha_{S}\beta_{D}}(b'-r_{D})^{\alpha_{D}\beta_{D}}}{(b'+r_{D}+r_{S})^{\alpha_{D}\beta_{M}}(b'-r_{S})^{\alpha_{D}\beta_{S}}(r_{S}-a')^{\alpha_{S}\beta_{S}}(a'+r_{D}+r_{S})^{\alpha_{S}\beta_{M}}}\Big] \nn\\
    &\, \sim \frac{1}{2}\,
    \frac{(b'-r_{D})^{\alpha_{D}\beta_{D}}}{(r_{S}-a')^{\alpha_{S}\beta_{S}}}\,{1\over C(a')}\ ,
\end{align}
using the approximation (\ref{b'-rD>>rs-a', b>>rs}), and (\ref{C(a')rDab}).
Thus we obtain
\begin{align}\label{log.a'+b'+a'-b'-/a'+b'-a'-b'+}
\frac{c}{3} \log\frac{d(a_{+},b_{+}')d(a_{-},b_{-}')}{d(a_{+},b_{-}')d(a_{-},b_{+}')}\,
  &\, =\,\frac{c}{3} \log\Big[1-2\frac{(r_{S}-a')^{\alpha_{S}\beta_{S}}}{(b'-r_{D})^{\alpha_{D}\beta_{D}}}\, C(a')\, \cosh{(X_{a'}+X_{b'})}\Big]\nn\\
&\quad\ -\frac{c}{3} \log\Big[1-2\frac{(r_{S}-a')^{\alpha_{S}\beta_{S}}}{(b'-r_{D})^{\alpha_{D}\beta_{D}}}\, C(a')\, \cosh{(X_{a'}-X_{b'})}\Big]\ .
\end{align}
The total bulk matter entanglement entropy thus is (\ref{log.a'+a'-b'+b'-})
plus (\ref{log.a'+b'+a'-b'-/a'+b'-a'-b'+}), along with the area term. Thus
at large values of $X_a'$ and $X_b'$, after adding the area term the total
entanglement entropy $S_{total}$ becomes
(\ref{Stotal=area+log.a'+a'-b'+b'-.a'+b'+a'-b'-/a'+b'-a'-b'+})\ \ie\
\begin{align}
    S_{total} & \sim\ \frac{2\pi {a'}^2}{G_{_N}} + \frac{2c}{6} \log\Big[\frac{2^2 \lambda}{\alpha_{D}\alpha_{S}}\sqrt{(r_{S}-a')(b'-r_{S})}\cdot \nn\\
    & \qquad\qquad
    \sqrt{\frac{(r_{D}-a')}{l}\frac{(b'-r_{D})}{l}\frac{(a'+r_{S}+r_{D})}{l}\frac{(b'+r_{S}+r_{D})}{l}} \sinh{X_{a'}} \sinh{X_{b'}}\Big]\nn\\
    & \qquad\qquad\qquad\qquad
    +\frac{c}{3} \log\Big[1-2\frac{(r_{S}-a')^{\alpha_{S}\beta_{S}}}{(b'-r_{D})^{\alpha_{D}\beta_{D}}}\, C(a')\, \cosh{(X_{a'}+X_{b'})}\Big]\nn\\
    & \qquad\qquad\qquad\qquad
    -\frac{c}{3} \log\Big[1-2\frac{(r_{S}-a')^{\alpha_{S}\beta_{S}}}{(b'-r_{D})^{\alpha_{D}\beta_{D}}}\, C(a')\, \cosh{(X_{a'}-X_{b'})}\Big]\ .
\end{align}

\section{Inconsistencies in other island solutions}

In this section, we briefly describe inconsistencies in other potential
island solutions.

\subsection{Island outside the black hole horizon}\label{wrong1}

We discuss a potential island with boundary just outside the black
hole horizon \ie\ in the static diamond, similar to the results in
\cite{Goswami:2022ylc} reviewed in
sec.~\ref{detailed calculation of previous paper}.
Here we use the static diamond Kruskal coordinates 
(\ref{SdS-Kruskal in SP 1}) redefined as $X_a= \alpha_S t_a$ and
$T_a= \alpha_S r^{\ast}_a$ for the location of the island boundary.
The calculation now gives the total generalized entropy as
\begin{align}\label{Stotal=area+log.a'+a'-b'+b'-.a'+b'+a'-b'-/a'+b'-a'-b'+(I outside horizon)}
    S_{total} & =\frac{2\pi {a}^{2}}{G_{_N}}+\frac{c}{6} \log \Big[\frac{2^4\lambda^{2}}{\alpha_{D}^2\alpha_{S}^2}(a-r_{S})(b'-r_{S})(\frac{r_{D}-a}{l})(\frac{b'-r_{D}}{l})
    \cdot \nn\\
    & \qquad\qquad\qquad\qquad
    (\frac{a+r_{S}+r_{D}}{l})(\frac{b+r_{S}+r_{D}}{l}) \cosh^2{X_{a}} \sinh^2{X_{b'}}\Big]\nn\\
    & \qquad\qquad\qquad + \frac{c}{3} \log \Big[1+2\frac{(a-r_{S})^{\alpha_{S}\beta_{S}}}{(b'-r_{D})^{\alpha_{D}\beta_{D}}}\,C(a)\, \sinh{(X_{a}+X_{b'})}\Big]\nn\\
    & \qquad\qquad\qquad -\frac{c}{3} \log \Big[1+2\frac{(a-r_{S})^{\alpha_{S}\beta_{S}}}{(b'-r_{D})^{\alpha_{D}\beta_{D}}}\,C(a)\, \sinh{(X_{a}-X_{b'})}\Big]\ ,
\end{align}
Extremizing (\ref{Stotal=area+log.a'+a'-b'+b'-.a'+b'+a'-b'-/a'+b'-a'-b'+(I outside horizon)}) with the island boundary $a$ as
$\frac{\partial S_{total}}{\partial a}=0$ gives
\begin{align}\label{simplified extremize (I outside horizon)}
&\frac{4 \pi a}{G_{_N}}+ \frac{c}{6}\frac{1}{a-r_S}+ \frac{c}{3}{\frac{C(a)}{\sqrt{b'-r_D}\sqrt{a-r_S}}}\, \cdot\Big[\frac{\sinh({X_{a}+X_{b'}})}{1+2\frac{(a-r_{S})^{\alpha_{S}\beta_{S}}}{(b'-r_{D})^{\alpha_{D}\beta_{D}}}\,C(a)\, \sinh{(X_{a}+X_{b'})}} \nn\\ 
&\qquad\qquad\qquad\qquad\qquad\qquad - \frac{\sinh({X_{a}-X_{b'}})}{1+2\frac{(a-r_{S})^{\alpha_{S}\beta_{S}}}{(b'-r_{D})^{\alpha_{D}\beta_{D}}}\,C(a)\, \sinh{(X_{a}-X_{b'})}}\Big] =0\ .
\end{align} 
Next, extremizing (\ref{Stotal=area+log.a'+a'-b'+b'-.a'+b'+a'-b'-/a'+b'-a'-b'+(I outside horizon)}) with respect to $X_{a}$ as $\frac{\partial S_{total}}{\partial X_{a}}=0$ gives
\begin{align}\label{extremize X_a'(I outside horizon)}
&\tanh{X_{a}}= 2 \sqrt{\frac{a-r_S}{b'-r_D}}\ C(a) \cdot \Big[\frac{\cosh({X_{a}-X_{b'}})}{1+2\frac{(a-r_{S})^{\alpha_{S}\beta_{S}}}{(b'-r_{D})^{\alpha_{D}\beta_{D}}}\,C(a)\, \sinh{(X_{a}-X_{b'})}} \nn\\
&\qquad\qquad\qquad\qquad\qquad\qquad\qquad
- \frac{\cosh({X_{a}+X_{b'}})}{1+2\frac{(a-r_{S})^{\alpha_{S}\beta_{S}}}{(b'-r_{D})^{\alpha_{D}\beta_{D}}}\,C(a)\, \sinh{(X_{a}+X_{b'})}}\Big]
\end{align}
Considering $X_a=X_{b'}$, (\ref{extremize X_a'(I outside horizon)}) becomes
\begin{align}\label{final X_a'=X_b'}
  1+2 \sqrt{\frac{a-r_S}{b'-r_{D}}} C(a)\sinh{2X_{a}}
  & =   2\sqrt{\frac{a-r_S}{b'-r_{D}}} C(a) \frac{1}{\tanh{X_{a}}} \nn \\&
 \cdot [1 +2 \sqrt{\frac{a-r_S}{b'-r_D}} C(a)\sinh{2X_{a}} - \cosh{2X_{a}}]\,.
\end{align}
Putting this condition (\ref{final X_a'=X_b'}) back in
(\ref{simplified extremize (I outside horizon)}) for large $X_{a}$
(with $a-r_S$ small) gives
\begin{equation}\label{final a'}
    a+ \frac{G_N c}{24 \pi} \frac{1}{a-r_{S}}\Big(1-\frac{\tanh{X_{a}}}{\coth{2X_{a}}}\Big)=0\,.
\end{equation}
It can be seen that $\frac{\tanh{X_{a}}}{\coth{2X_{a}}}<1$ always so that
all terms are positive here: thus there is no solution with $a>r_S$.
Thus these extremization equations
(\ref{simplified extremize (I outside horizon)}) and
(\ref{extremize X_a'(I outside horizon)}) together do not give
reasonable island solutions.

Similarly, if we consider $X_a=-X_{b'}$, (\ref{extremize X_a'(I outside horizon)}) becomes
\begin{align}\label{final X_a'=-X_b'}
  1+2 \sqrt{\frac{a-r_S}{b'-r_{D}}} C(a)\sinh{2X_{a}}
 & =   2\sqrt{\frac{a-r_S}{b'-r_{D}}} C(a) \frac{1}{\tanh{X_{a}}} \nn \\&
  \cdot [- 1 - 2 \sqrt{\frac{a-r_S}{b'-r_D}} C(a)\sinh{2X_{a}} + \cosh{2X_{a}}]\,.
\end{align}
Putting this condition (\ref{final X_a'=-X_b'}) back in
(\ref{simplified extremize (I outside horizon)}) for large $X_{a}$
gives (\ref{final a'}) again.

\subsection{Island inside the black hole: another possibility}\label{wrong2}

Recalling sec.~\ref{sec:EEisland} and the extremization equations
(\ref{simplified extremize}),\ (\ref{extremize X_a'}).
Instead of $X_{a'}=X_{b'}$ considered there, let us consider
$X_{a'}=-X_{b'}$: then $X_{a'}+X_{b'}=0$ and $X_{a'}-X_{b'}=2X_{a'}$.
Then (\ref{extremize X_a'}) gives
for large $X_{a'}$ and $X_{b'}$:
\begin{equation}\label{final X_a' eq 2}
1-2 \sqrt{\frac{r_S-a'}{b'-r_{D}}} C(a')\cosh{2X_{a'}}= -2\sqrt{\frac{r_S-a'}{b'-r_{D}}} C(a') \frac{\sinh{2X_{a'}}}{\coth{X_{a'}}}\,.
\end{equation}
The minus sign on the right hand side leads to trouble when this
is put back in (\ref{simplified extremize}), giving no semiclassical
$a'\lesssim r_S$ near horizon island solution.

\section{Future boundary, timelike separated QES}\label{timelike separated qes}

In this section, we exhibit other quantum extremal surface solutions
which are timelike separated from the radiation region near the future
boundary. We will use several technical details from
\cite{Fernandes:2019ige}.

The Schwarzschild de Sitter metric (\ref{SdSst}), 
after the redefinitions\ $\tau = \frac{l}{r},\ \omega = \frac{t}{l}$,
becomes 
\be\label{sch de sitter metric in new co-ordinate}
ds^2=\frac{l^2}{\tau^2} (-\frac{d\tau^2}{f(\tau)}+f(\tau)d\omega^2+d\Omega_{2}^2) ,  \quad\ \  f(\tau)= 1-\tau^2+ \frac{2m}{l}\tau^{3}=(1-a_{1}\tau)(1-a_{2}\tau)(1+(a_{1}+a_{2})\tau), \nn \\
\ee
\be\label{constraint}
a_{1}a_{2}(a_{1}+a_{2})=\frac{2m}{l}\,,\qquad  a_{1}^2+a_{1}a_{2}+a_{2}^2=1; \qquad 0 < a_{2} < a_{1} < 1\ ;\qquad {\frac{m}{l}}\leq {\frac{1}{3\sqrt{3}} } .
\ee
In the above, $\tau_{c}=\frac{1}{a_{1}}$ and $\tau_{s}=\frac{1}{a_{2}}$ are
the cosmological (de Sitter) and Schwarzschild horizons. (The third zero
does not correspond to a physical horizon.)

For $SdS_{4}$ with $f(\tau)$ in (\ref{sch de sitter metric in new co-ordinate}), the tortoise coordinate $y=\int \frac{d\tau}{f(\tau)}$ can be defined as
\be\label{sds tortoise}
y=\int \frac{d\tau}{1-\tau^2+ \frac{2m}{l}\tau^{3}}= - \beta_{1} \log(1-a_{1}\tau)+ \beta_{2} \log(1-a_{2}\tau) + \beta_{3} \log(1+(a_{1}+a_{2})\tau)\,.
\ee
With the parameters 
\be\label{parameters}
\beta_{1}=\frac{a_{1}}{3a_{1}^{2}-1}, \qquad \beta_{2}=-\frac{a_{2}}{3a_{2}^{2}-1}, \qquad \beta_{3}=\frac{a_{1}+a_{2}}{3a_{1}a_{2}+2}
\ee
the $SdS_4$ metric becomes
\be\label{sds4 metric with tortoise}
ds^2=l^2(\frac{1}{\tau^2}-1+\frac{2m}{l}\tau)(d\omega^2-dy^2)+\frac{l^2}{\tau^2}d\Omega_{2}^2\,.
\ee
Now, we consider the same reduction ansatz from \cite{Goswami:2022ylc}
to perform dimentional reduction of the $SdS_4$ background to
2-dimensions. The 2-dim metric and dilaton become
\be\label{sch de sitter metric in 2d}
ds^2_{2}=\frac{\lambda l^3}{\tau} (\frac{1}{\tau^2}-1+\frac{2m}{l}\tau)(d\omega^2-dy^2), \qquad \phi = \phi_{r}\frac{\lambda^{2}l^{2}}{\tau^{2}}\,.
\ee
In the reduced $2$-dim Schwarzschild de sitter spacetime,the Kruskal
coordinates around the cosmological horizon are then defined as $U, V$
and the metric becomes
\bea\label{sch de sitter metric in kruskal}
U=e^{\frac{\omega-y}{2 \beta_{1}}},\quad V=-e^{-\frac{\omega+y}{2 \beta_{1}}};
\qquad UV=-e^{-\frac{y}{\beta_{1}}},\quad \frac{U}{V}=-e^{\frac{\omega}{\beta_{1}}};
\nn\\ [1mm] 
UV= (a_{1}\tau-1)(1-a_{2}\tau)^{-\frac{\beta_{2}}{\beta_{1}}}(1+(a_{1}+a_{2})\tau)^{-\frac{\beta_{3}}{\beta_{1}}}\,,
\eea
\be\label{sch meric in U,V}
 ds_{2}^2=\frac{4\beta_{1}^{2} g(U,V)}{UV}\,dUdV\,.
 \ee
Using (\ref{sch de sitter metric in 2d}), the generalized entropy for
a future boundary observer at $(\omega, \tau)=(\omega_{0}, \tau_{0})$
in the $2$-dim $SdS_4$ spacetime becomes (with
$P(\tau_{0})=\frac{1}{\tau_{0}^{2}}-1+\frac{2m}{l}\tau_{0}$)
\be\label{S_gen sch de sitter}
S_{gen}=\frac{\phi_{r}}{4G} \frac{l^{2}}{\tau^{2}} + \frac{c}{12} \log[\frac{1}{\epsilon_{uv}^{4}} ((\omega-\omega_{0})^{2}-(y-y_{0})^{2})^{2}\frac{\lambda^{2}l^{6}}{\tau \tau_{0}}P(\tau_{0})(\frac{1}{\tau^{2}}-1+\frac{2m}{l}\tau)]\,.
\ee
This expression for $S_{gen}$ should be regarded as a smooth function
of $U, V$, with respect to which we will extremize to find quantum
extremal surfaces. However the nature of the QES here can be gleaned
by simply noting that the only place the spatial future boundary
coordinate $\omega$ enters is through the spacetime interval
$\Delta^{2}=(\omega-\omega_{0})^{2}-(y-y_{0})^{2}$ inside the
logarithm. Thus we expect\
$\frac{\partial S_{gen}}{\partial \omega}
= \frac{c}{3} \frac{\omega-\omega_{0}}{\Delta^{2}} = 0$,\
so that $\omega=\omega_0$, \ie\ the QES is timelike separated from the
future boundary observer.

Analysing more carefully, extremizing (\ref{S_gen sch de sitter})
with Kruskal $U$, $V$, as
\be\label{ext U2}
\frac{\partial S_{gen}}{\partial U}= \frac{\partial S_{gen}}{\partial \tau} \frac{\partial \tau}{\partial U} + \frac{\partial S_{gen}}{\partial \omega} \frac{\partial \omega}{\partial U} =0 \qquad, \qquad
\frac{\partial S_{gen}}{\partial V}= \frac{\partial S_{gen}}{\partial \tau} \frac{\partial \tau}{\partial V} + \frac{\partial S_{gen}}{\partial \omega} \frac{\partial \omega}{\partial V} =0\,,
\ee
and, from (\ref{S_gen sch de sitter}),
(\ref{sch de sitter metric in kruskal}), (\ref{sch meric in U,V}), noting that
\be\label{ext tau,omega sds}
\frac{\partial S_{gen}}{\partial \tau}= - \frac{\phi_{r}}{2G} \frac{l^{2}}{\tau^{3}} + \frac{c}{12} \frac{1}{f(\tau)} [- \frac{4(y-y_{0})}{\Delta^{2}}+ \tau(1-\frac{3}{\tau^{2}})], \qquad \frac{\partial S_{gen}}{\partial \omega}= \frac{c}{3} \frac{\omega-\omega_{0}}{\Delta^{2}}; \nn \\
\ee 
we find the extremization conditions become
\be\label{final ext U2}
[- \frac{\phi_{r}}{2G} \frac{l^{2}}{\tau^{3}} + \frac{c}{12} \frac{1}{f(\tau)} (- \frac{4(y-y_{0})}{\Delta^{2}}+ \tau(1-\frac{3}{\tau^{2}}))] \frac{A(\tau)}{U} +
\frac{c}{3} \frac{\omega-\omega_{0}}{\Delta^{2}} \frac{\beta_{1}}{U} =0\,,
\ee
and
\be\label{final ext V2}
[- \frac{\phi_{r}}{2G} \frac{l^{2}}{\tau^{3}} + \frac{c}{12} \frac{1}{f(\tau)} (- \frac{4(y-y_{0})}{\Delta^{2}}+ \tau(1-\frac{3}{\tau^{2}}))] \frac{A(\tau)}{V} +
\frac{c}{3} \frac{\omega-\omega_{0}}{\Delta^{2}} (-\frac{\beta_{1}}{V}) =0\,.
\ee
Subtracting as (\ref{final ext U2})-(\ref{final ext V2})
gives $\frac{2c}{3} \frac{\omega-\omega_{0}}{\Delta^{2}} \beta_{1}=0 $.\
Thus,
\be \label{omega2 eq}
\omega=\omega_{0}\,,
\ee
giving timelike separated QES with respect to the future boundary observer.
The timelike separation implies that the generalized entropy becomes
complex-valued, with $\log(-|\Delta^2|)$ giving rise to $\log(-1)=i\pi$,
as in pure de Sitter.

Now, by putting $\omega=\omega_{0}$ in (\ref{final ext U2}) gives
\be\label{y1 eq}
- \frac{\phi_{r}}{2G} \frac{l^{2}}{\tau^{3}} + \frac{c}{12} \frac{1}{f(\tau)} ( \frac{4}{y-y_{0}}+ \tau(1-\frac{3}{\tau^{2}}))=0
\ee
The future boundary observer has $r=\infty$ and $\tau_{0}=0$ so
$y_{0}=0 $. Further considering $\tau=1-\epsilon$, where $\epsilon \ll 1$,
and noting  $1 \ll c \ll \frac{1}{G}$, we simplify (\ref{y1 eq})
ignoring appropriate terms and obtain
\be\label{tau3 eq}
\frac{2c}{\log[\frac{(1-a_{2}(1-\epsilon))^{\beta_{2}}(1+(a_{1}+a_{2})(1-\epsilon))^{\beta_{3}}}{(1-a_{1}(1-\epsilon))^{\beta_{1}}}]} \approx \frac{3\phi_{r}}{G} l^{2}[\frac{\epsilon(2-\epsilon)}{(1-\epsilon)^{3}}+\frac{2m}{l}] 
\ee 
Now, considering $a_{1}$ and $a_{2}$ perturbatively as
$a_{1}\simeq 1-\frac{m}{l}$,\ $a_{2} \simeq \frac{2m}{l}$, and
$0 < a_{2} < a_{1} < 1$, we can obtain the parameters $\beta_{1},\beta_{2},\beta_{3}$ using (\ref{parameters}) perturbatively as well. This finally gives
\be\label{c condition sch de sitter}
c \approx \frac{3\phi_{r}}{2G}l^{2} (\epsilon+\frac{m}{l}) \log(\frac{2}{\epsilon+\frac{m}{l}})
\ee
which is a consistency condition on the central charge (number of degrees
of freedom) of the 2-dim CFT matter for the timelike extremal surface
to exist (pure $dS$ corresponds to $m=0$).


\begin{thebibliography}{} 

{ \renewcommand{\baselinestretch}{0.5}
\scriptsize{

\bibitem{Hawking:1976ra}
S.~W.~Hawking,
``Breakdown of Predictability in Gravitational Collapse,''
Phys. Rev. D \textbf{14}, 2460-2473 (1976)
doi:10.1103/PhysRevD.14.2460

\bibitem{Hawking:1975vcx}
S.~W.~Hawking,
``Particle Creation by Black Holes,''
Commun. Math. Phys. \textbf{43}, 199-220 (1975)
[erratum: Commun. Math. Phys. \textbf{46}, 206 (1976)]
doi:10.1007/BF02345020

\bibitem{Mathur:2009hf}
S.~D.~Mathur,
``The Information paradox: A Pedagogical introduction,''
Class. Quant. Grav. \textbf{26}, 224001 (2009)
doi:10.1088/0264-9381/26/22/224001
[arXiv:0909.1038 [hep-th]].

\bibitem{Almheiri:2012rt}
A.~Almheiri, D.~Marolf, J.~Polchinski and J.~Sully,
``Black Holes: Complementarity or Firewalls?,''
JHEP \textbf{02}, 062 (2013)
doi:10.1007/JHEP02(2013)062
[arXiv:1207.3123 [hep-th]].

\bibitem{Page:1993wv}
D.~N.~Page,
``Information in black hole radiation,''
Phys. Rev. Lett. \textbf{71}, 3743-3746 (1993)
doi:10.1103/PhysRevLett.71.3743
[arXiv:hep-th/9306083 [hep-th]].

\bibitem{Page:2013dx}
D.~N.~Page,
``Time Dependence of Hawking Radiation Entropy,''
JCAP \textbf{09}, 028 (2013)
doi:10.1088/1475-7516/2013/09/028
[arXiv:1301.4995 [hep-th]].

\bibitem{Penington:2019npb}
G.~Penington,
``Entanglement Wedge Reconstruction and the Information Paradox,''
JHEP \textbf{09}, 002 (2020)
doi:10.1007/JHEP09(2020)002
[arXiv:1905.08255 [hep-th]].

\bibitem{Almheiri:2019psf}
A.~Almheiri, N.~Engelhardt, D.~Marolf and H.~Maxfield,
``The entropy of bulk quantum fields and the entanglement wedge of an evaporating black hole,''
JHEP \textbf{12}, 063 (2019)
doi:10.1007/JHEP12(2019)063
[arXiv:1905.08762 [hep-th]].

\bibitem{Almheiri:2019hni}
A.~Almheiri, R.~Mahajan, J.~Maldacena and Y.~Zhao,
``The Page curve of Hawking radiation from semiclassical geometry,''
JHEP \textbf{03}, 149 (2020)
doi:10.1007/JHEP03(2020)149
[arXiv:1908.10996 [hep-th]].

\bibitem{Penington:2019kki}
G.~Penington, S.~H.~Shenker, D.~Stanford, Z.~Yang,
``Replica wormholes \& the black hole interior,''
[arXiv:1911.11977[hep-th]].

\bibitem{Almheiri:2019qdq}
A.~Almheiri, T.~Hartman, J.~Maldacena, E.~Shaghoulian and A.~Tajdini,
``Replica Wormholes and the Entropy of Hawking Radiation,''
JHEP \textbf{05}, 013 (2020)
[arXiv:1911.12333 [hep-th]].

\bibitem{Faulkner:2013ana} 
  T.~Faulkner, A.~Lewkowycz and J.~Maldacena,
  ``Quantum corrections to holographic entanglement entropy,''
  JHEP {\bf 1311}, 074 (2013)
  doi:10.1007/JHEP11(2013)074
  [arXiv:1307.2892 [hep-th]].

\bibitem{Engelhardt:2014gca}
N.~Engelhardt and A.~C.~Wall,
``Quantum Extremal Surfaces: Holographic Entanglement Entropy beyond the Classical Regime,''
JHEP \textbf{01}, 073 (2015)
doi:10.1007/JHEP01(2015)073
[arXiv:1408.3203 [hep-th]].

\bibitem{Ryu:2006bv} 
  S.~Ryu and T.~Takayanagi,
  ``Holographic derivation of entanglement entropy from AdS/CFT,''
  Phys.\ Rev.\ Lett.\  {\bf 96}, 181602 (2006)
  [hep-th/0603001].

\bibitem{Ryu:2006ef} 
  S.~Ryu and T.~Takayanagi,
  ``Aspects of Holographic Entanglement Entropy,''
  JHEP {\bf 0608}, 045 (2006)
  [hep-th/0605073].

\bibitem{HRT} 
V.~E.~Hubeny, M.~Rangamani and T.~Takayanagi,
``A Covariant holographic entanglement entropy proposal,'' 
JHEP {\bf 0707} (2007) 062  [arXiv:0705.0016 [hep-th]].

\bibitem{Rangamani:2016dms} 
  M.~Rangamani and T.~Takayanagi,
  ``Holographic Entanglement Entropy,''
  Lect.\ Notes Phys.\  {\bf 931}, pp.1 (2017)
  [arXiv:1609.01287 [hep-th]].

\bibitem{Almheiri:2020cfm}
A.~Almheiri, T.~Hartman, J.~Maldacena, E.~Shaghoulian and A.~Tajdini,
``The entropy of Hawking radiation,''
[arXiv:2006.06872 [hep-th]].

\bibitem{Raju:2020smc}
S.~Raju,
``Lessons from the Information Paradox,''
[arXiv:2012.05770 [hep-th]].

\bibitem{Chen:2021lnq}
B.~Chen, B.~Czech and Z.~z.~Wang,
``Quantum Information in Holographic Duality,''
[arXiv:2108.09188 [hep-th]].

\bibitem{Almheiri:2019yqk}
A.~Almheiri, R.~Mahajan and J.~Maldacena,
``Islands outside the horizon,''
[arXiv:1910.11077 [hep-th]].

\bibitem{Chen:2019uhq}
H.~Z.~Chen, Z.~Fisher, J.~Hernandez, R.~C.~Myers and S.~M.~Ruan,
``Information Flow in Black Hole Evaporation,''
JHEP \textbf{03}, 152 (2020)
doi:10.1007/JHEP03(2020)152
[arXiv:1911.03402 [hep-th]].

\bibitem{Almheiri:2019psy}
A.~Almheiri, R.~Mahajan and J.~E.~Santos,
``Entanglement islands in higher dimensions,''
SciPost Phys. \textbf{9}, no.1, 001 (2020)
doi:10.21468/SciPostPhys.9.1.001
[arXiv:1911.09666 [hep-th]].

\bibitem{Gautason:2020tmk}
F.~F.~Gautason, L.~Schneiderbauer, W.~Sybesma and L.~Thorlacius,
``Page Curve for an Evaporating Black Hole,''
JHEP \textbf{05}, 091 (2020)
doi:10.1007/JHEP05(2020)091
[arXiv:2004.00598 [hep-th]].

\bibitem{Anegawa:2020ezn}
T.~Anegawa and N.~Iizuka,
``Notes on islands in asymptotically flat 2d dilaton black holes,''
JHEP \textbf{07}, 036 (2020)
doi:10.1007/JHEP07(2020)036
[arXiv:2004.01601 [hep-th]].

\bibitem{Hashimoto:2020cas}
K.~Hashimoto, N.~Iizuka and Y.~Matsuo,
``Islands in Schwarzschild black holes,''
JHEP \textbf{06}, 085 (2020)
doi:10.1007/JHEP06(2020)085
[arXiv:2004.05863 [hep-th]].

\bibitem{Hartman:2020swn}
T.~Hartman, E.~Shaghoulian and A.~Strominger,
``Islands in Asymptotically Flat 2D Gravity,''
JHEP \textbf{07}, 022 (2020)
doi:10.1007/JHEP07(2020)022
[arXiv:2004.13857 [hep-th]].

\bibitem{Hollowood:2020cou}
T.~J.~Hollowood and S.~P.~Kumar,
``Islands and Page Curves for Evaporating Black Holes in JT Gravity,''
JHEP \textbf{08}, 094 (2020)
doi:10.1007/JHEP08(2020)094
[arXiv:2004.14944 [hep-th]].

\bibitem{Krishnan:2020oun}
C.~Krishnan, V.~Patil and J.~Pereira,
``Page Curve and the Information Paradox in Flat Space,''
[arXiv:2005.02993 [hep-th]].

\bibitem{Alishahiha:2020qza}
M.~Alishahiha, A.~Faraji Astaneh and A.~Naseh,
``Island in the presence of higher derivative terms,''
JHEP \textbf{02}, 035 (2021)
doi:10.1007/JHEP02(2021)035
[arXiv:2005.08715 [hep-th]].

\bibitem{Geng:2020qvw}
H.~Geng and A.~Karch,
``Massive islands,''
JHEP \textbf{09}, 121 (2020)
doi:10.1007/JHEP09(2020)121
[arXiv:2006.02438 [hep-th]].

\bibitem{Li:2020ceg}
T.~Li, J.~Chu and Y.~Zhou,
``Reflected Entropy for an Evaporating Black Hole,''
JHEP \textbf{11}, 155 (2020)
doi:10.1007/JHEP11(2020)155
[arXiv:2006.10846 [hep-th]].


\bibitem{Dong:2020uxp}
X.~Dong, X.~L.~Qi, Z.~Shangnan and Z.~Yang,
``Effective entropy of quantum fields coupled with gravity,''
JHEP \textbf{10}, 052 (2020)
doi:10.1007/JHEP10(2020)052
[arXiv:2007.02987 [hep-th]].

\bibitem{Chen:2020jvn}
H.~Z.~Chen, Z.~Fisher, J.~Hernandez, R.~C.~Myers and S.~M.~Ruan,
``Evaporating Black Holes Coupled to a Thermal Bath,''
JHEP \textbf{01}, 065 (2021)
doi:10.1007/JHEP01(2021)065
[arXiv:2007.11658 [hep-th]].

\bibitem{Ling:2020laa}
Y.~Ling, Y.~Liu and Z.~Y.~Xian,
``Island in Charged Black Holes,''
JHEP \textbf{03}, 251 (2021)
doi:10.1007/JHEP03(2021)251
[arXiv:2010.00037 [hep-th]].

\bibitem{Matsuo:2020ypv}
Y.~Matsuo,
``Islands and stretched horizon,''
JHEP \textbf{07}, 051 (2021)
[arXiv:2011.08814 [hep-th]].

\bibitem{Goto:2020wnk}
K.~Goto, T.~Hartman and A.~Tajdini,
``Replica wormholes for an evaporating 2D black hole,''
JHEP \textbf{04}, 289 (2021)
doi:10.1007/JHEP04(2021)289
[arXiv:2011.09043 [hep-th]].

\bibitem{Akal:2020twv}
I.~Akal, Y.~Kusuki, N.~Shiba, T.~Takayanagi and Z.~Wei,
``Entanglement Entropy in a Holographic Moving Mirror and the Page Curve,''
Phys. Rev. Lett. \textbf{126}, no.6, 061604 (2021)
doi:10.1103/PhysRevLett.126.061604
[arXiv:2011.12005 [hep-th]].

\bibitem{Geng:2020fxl}
H.~Geng, A.~Karch, C.~Perez-Pardavila, S.~Raju, L.~Randall, M.~Riojas and S.~Shashi,
``Information Transfer with a Gravitating Bath,''
SciPost Phys. \textbf{10}, no.5, 103 (2021)
doi:10.21468/SciPostPhys.10.5.103
[arXiv:2012.04671 [hep-th]].

\bibitem{Deng:2020ent}
F.~Deng, J.~Chu and Y.~Zhou,
``Defect extremal surface as the holographic counterpart of Island formula,''
JHEP \textbf{03}, 008 (2021)
doi:10.1007/JHEP03(2021)008
[arXiv:2012.07612 [hep-th]].

\bibitem{Karananas:2020fwx}
G.~K.~Karananas, A.~Kehagias and J.~Taskas,
``Islands in linear dilaton black holes,''
JHEP \textbf{03}, 253 (2021)
doi:10.1007/JHEP03(2021)253
[arXiv:2101.00024 [hep-th]].

\bibitem{Wang:2021woy}
X.~Wang, R.~Li and J.~Wang,
``Islands and Page curves of Reissner-Nordstr\"om black holes,''
JHEP \textbf{04}, 103 (2021)
doi:10.1007/JHEP04(2021)103
[arXiv:2101.06867 [hep-th]].

\bibitem{Verheijden:2021yrb}
E.~Verheijden and E.~Verlinde,
``From the BTZ black hole to JT gravity: geometrizing the island,''
JHEP \textbf{11}, 092 (2021)
doi:10.1007/JHEP11(2021)092
[arXiv:2102.00922 [hep-th]].

\bibitem{Kawabata:2021hac}
K.~Kawabata, T.~Nishioka, Y.~Okuyama and K.~Watanabe,
``Probing Hawking radiation through capacity of entanglement,''
JHEP \textbf{05}, 062 (2021)
doi:10.1007/JHEP05(2021)062
[arXiv:2102.02425 [hep-th]].

\bibitem{Anderson:2020vwi}
L.~Anderson, O.~Parrikar and R.~M.~Soni,
``Islands with gravitating baths: towards ER = EPR,''
JHEP \textbf{21}, 226 (2020)
doi:10.1007/JHEP10(2021)226
[arXiv:2103.14746 [hep-th]].

\bibitem{Bhattacharya:2021jrn}
A.~Bhattacharya, A.~Bhattacharyya, P.~Nandy and A.~K.~Patra,
``Islands and complexity of eternal black hole and radiation subsystems for a doubly holographic model,''
JHEP \textbf{05}, 135 (2021)
doi:10.1007/JHEP05(2021)135
[arXiv:2103.15852 [hep-th]].

\bibitem{Kim:2021gzd}
W.~Kim and M.~Nam,
``Entanglement entropy of asymptotically flat non-extremal and extremal black holes with an island,''
Eur. Phys. J. C \textbf{81}, no.10, 869 (2021)
doi:10.1140/epjc/s10052-021-09680-x
[arXiv:2103.16163 [hep-th]].

\bibitem{Ghosh:2021axl}
K.~Ghosh and C.~Krishnan,
``Dirichlet baths and the not-so-fine-grained Page curve,''
JHEP \textbf{08}, 119 (2021)
doi:10.1007/JHEP08(2021)119
[arXiv:2103.17253 [hep-th]].

\bibitem{Wang:2021mqq}
X.~Wang, R.~Li and J.~Wang,
``Page curves for a family of exactly solvable evaporating black holes,''
Phys. Rev. D \textbf{103}, no.12, 126026 (2021)
doi:10.1103/PhysRevD.103.126026
[arXiv:2104.00224 [hep-th]].

\bibitem{Li:2021lfo}
R.~Li, X.~Wang and J.~Wang,
``Island may not save the information paradox of Liouville black holes,''
Phys. Rev. D \textbf{104}, no.10, 106015 (2021)
doi:10.1103/PhysRevD.104.106015
[arXiv:2105.03271 [hep-th]].

\bibitem{Li:2021mjp}
R.~Li and J.~Wang,
``Hawking radiation and page curves of the black holes in thermal environment,''
Commun. Theor. Phys. \textbf{73}, no.7, 075401 (2021)
doi:10.1088/1572-9494/abf823

\bibitem{Kawabata:2021vyo}
K.~Kawabata, T.~Nishioka, Y.~Okuyama and K.~Watanabe,
``Replica wormholes and capacity of entanglement,''
JHEP \textbf{10}, 227 (2021)
doi:10.1007/JHEP10(2021)227
[arXiv:2105.08396 [hep-th]].

\bibitem{Lu:2021gmv}
Y.~Lu, J.~Lin,
``Islands in Kaluza\textendash{}Klein black holes,''
Eur. Phys. J. C \textbf{82}, no.2, 132 (2022)
[arXiv:2106.07845 [hep-th]].

\bibitem{Kruthoff:2021vgv}
J.~Kruthoff, R.~Mahajan and C.~Murdia,
``Free fermion entanglement with a semitransparent interface: the effect of graybody factors on entanglement islands,''
SciPost Phys. \textbf{11}, 063 (2021)
[arXiv:2106.10287 [hep-th]].

\bibitem{Yu:2021cgi}
M.~H.~Yu and X.~H.~Ge,
``Islands and Page curves in charged dilaton black holes,''
Eur. Phys. J. C \textbf{82}, no.1, 14 (2022)
doi:10.1140/epjc/s10052-021-09932-w
[arXiv:2107.03031 [hep-th]].

\bibitem{Ahn:2021chg}
B.~Ahn, S.~E.~Bak, H.~S.~Jeong, K.~Y.~Kim and Y.~W.~Sun,
``Islands in charged linear dilaton black holes,''
Phys. Rev. D \textbf{105}, no.4, 046012 (2022)
doi:10.1103/PhysRevD.105.046012
[arXiv:2107.07444 [hep-th]].

\bibitem{Wang:2021afl}
X.~Wang, K.~Zhang and J.~Wang,
``What can we learn about islands and state paradox from quantum information theory?,''
[arXiv:2107.09228 [hep-th]].

\bibitem{Arefeva:2021kfx}
I.~Aref'eva and I.~Volovich,
``A Note on Islands in Schwarzschild Black Holes,''
[arXiv:2110.04233 [hep-th]].


\bibitem{He:2021mst}
S.~He, Y.~Sun, L.~Zhao and Y.~X.~Zhang,
``The universality of islands outside the horizon,''
JHEP \textbf{05}, 047 (2022)
doi:10.1007/JHEP05(2022)047
[arXiv:2110.07598 [hep-th]].

\bibitem{Matsuo:2021mmi}
Y.~Matsuo,
``Entanglement entropy and vacuum states in Schwarzschild geometry,''
JHEP \textbf{06}, 109 (2022)
doi:10.1007/JHEP06(2022)109
[arXiv:2110.13898 [hep-th]].

\bibitem{Omidi:2021opl}
F.~Omidi,
``Entropy of Hawking radiation for two-sided hyperscaling violating black branes,''
JHEP \textbf{04}, 022 (2022)
doi:10.1007/JHEP04(2022)022
[arXiv:2112.05890 [hep-th]].

\bibitem{Espindola:2022fqb}
R.~Esp\'\i{}ndola, B.~Najian and D.~Nikolakopoulou,
``Islands in FRW Cosmologies,''
[arXiv:2203.04433 [hep-th]].

\bibitem{Tian:2022pso}
J.~Tian,
``Islands in Generalized Dilaton Theories,''
[arXiv:2204.08751 [hep-th]].

\bibitem{Laddha:2020kvp}
A.~Laddha, S.~G.~Prabhu, S.~Raju and P.~Shrivastava,
``The Holographic Nature of Null Infinity,''
SciPost Phys. \textbf{10}, no.2, 041 (2021)
doi:10.21468/SciPostPhys.10.2.041
[arXiv:2002.02448 [hep-th]].

\bibitem{Geng:2021hlu}
H.~Geng, A.~Karch, C.~Perez-Pardavila, S.~Raju, L.~Randall, M.~Riojas and S.~Shashi,
``Inconsistency of Islands in Theories with Long-Range Gravity,''
[arXiv:2107.03390 [hep-th]].

\bibitem{Bena:2022rna}
I.~Bena, E.~J.~Martinec, S.~D.~Mathur and N.~P.~Warner,
``Fuzzballs and Microstate Geometries: Black-Hole Structure in String Theory,''
[arXiv:2204.13113 [hep-th]].

\bibitem{Krishnan:2020fer}
C.~Krishnan,
``Critical Islands,''
JHEP \textbf{01}, 179 (2021)
doi:10.1007/JHEP01(2021)179
[arXiv:2007.06551 [hep-th]].

\bibitem{Chen:2020tes}
Y.~Chen, V.~Gorbenko, J.~Maldacena,
``Bra-ket wormholes in gravitationally prepared states,''
[arXiv:2007.16091 [hep-th]].

\bibitem{Hartman:2020khs}
T.~Hartman, Y.~Jiang and E.~Shaghoulian,
``Islands in cosmology,''
JHEP \textbf{11}, 111 (2020)
doi:10.1007/JHEP11(2020)111
[arXiv:2008.01022 [hep-th]].

\bibitem{VanRaamsdonk:2020tlr}
M.~Van Raamsdonk,
``Comments on wormholes, ensembles, and cosmology,''
arXiv:2008.02259[hep-th].

\bibitem{Balasubramanian:2020xqf}
V.~Balasubramanian, A.~Kar and T.~Ugajin,
``Islands in de Sitter space,''
JHEP \textbf{02}, 072 (2021)
doi:10.1007/JHEP02(2021)072
[arXiv:2008.05275 [hep-th]].

\bibitem{Sybesma:2020fxg}
W.~Sybesma,
``Pure de Sitter space and the island moving back in time,''
Class. Quant. Grav. \textbf{38}, no.14, 145012 (2021)
doi:10.1088/1361-6382/abff9a
[arXiv:2008.07994 [hep-th]].

\bibitem{Manu:2020tty}
A.~Manu, K.~Narayan and P.~Paul,
``Cosmological singularities, entanglement and quantum extremal surfaces,''
JHEP \textbf{04}, 200 (2021)
doi:10.1007/JHEP04(2021)200
[arXiv:2012.07351 [hep-th]].

\bibitem{Choudhury:2020hil}
S.~Choudhury, S.~Chowdhury, N.~Gupta, A.~Mishara, S.~P.~Selvam, S.~Panda, G.~D.~Pasquino, C.~Singha and A.~Swain,
``Circuit Complexity From Cosmological Islands,''
Symmetry \textbf{13}, 1301 (2021)
[arXiv:2012.10234 [hep-th]].

\bibitem{Bousso:2021sji}
R.~Bousso and A.~Shahbazi-Moghaddam,
``Island Finder and Entropy Bound,''
Phys. Rev. D \textbf{103}, no.10, 106005 (2021)
doi:10.1103/PhysRevD.103.106005
[arXiv:2101.11648 [hep-th]].

\bibitem{Geng:2021wcq}
H.~Geng, Y.~Nomura and H.~Y.~Sun,
``Information paradox and its resolution in de Sitter holography,''
Phys. Rev. D \textbf{103}, no.12, 126004 (2021)
doi:10.1103/PhysRevD.103.126004
[arXiv:2103.07477 [hep-th]].

\bibitem{Fallows:2021sge}
S.~Fallows and S.~F.~Ross,
``Islands and mixed states in closed universes,''
JHEP \textbf{07}, 022 (2021)
doi:10.1007/JHEP07(2021)022
[arXiv:2103.14364 [hep-th]].

\bibitem{Aalsma:2021bit}
L.~Aalsma and W.~Sybesma,
``The Price of Curiosity: Information Recovery in de Sitter Space,''
JHEP \textbf{05}, 291 (2021)
doi:10.1007/JHEP05(2021)291
[arXiv:2104.00006 [hep-th]].

\bibitem{Uhlemann:2021nhu}
C.~F.~Uhlemann,
``Islands and Page curves in 4d from Type IIB,''
JHEP \textbf{08}, 104 (2021)
doi:10.1007/JHEP08(2021)104
[arXiv:2105.00008 [hep-th]].

\bibitem{Giataganas:2021cwg}
D.~Giataganas and N.~Tetradis,
``Entanglement entropy in FRW backgrounds,''
Phys. Lett. B \textbf{820}, 136493 (2021)
doi:10.1016/j.physletb.2021.136493
[arXiv:2105.12614 [hep-th]].

\bibitem{Aalsma:2021kle}
L.~Aalsma, A.~Cole, E.~Morvan, J.~P.~van der Schaar and G.~Shiu,
``Shocks and information exchange in de Sitter space,''
JHEP \textbf{10}, 104 (2021)
doi:10.1007/JHEP10(2021)104
[arXiv:2105.12737 [hep-th]].

\bibitem{Langhoff:2021uct}
K.~Langhoff, C.~Murdia and Y.~Nomura,
``Multiverse in an inverted island,''
Phys. Rev. D \textbf{104}, no.8, 086007 (2021)
doi:10.1103/PhysRevD.104.086007
[arXiv:2106.05271 [hep-th]].

\bibitem{Aguilar-Gutierrez:2021bns}
S.~E.~Aguilar-Gutierrez, A.~Chatwin-Davies, T.~Hertog, N.~Pinzani-Fokeeva and B.~Robinson,
``Islands in Multiverse Models,''
[arXiv:2108.01278 [hep-th]].

\bibitem{Shaghoulian:2021cef}
E.~Shaghoulian,
``The central dogma and cosmological horizons,''
JHEP \textbf{01}, 132 (2022)
[arXiv:2110.13210 [hep-th]].

\bibitem{Chou:2021boq}
C.~J.~Chou, H.~B.~Lao and Y.~Yang,
``Page curve of effective Hawking radiation,''
Phys. Rev. D \textbf{106}, no.6, 066008 (2022)
doi:10.1103/PhysRevD.106.066008
[arXiv:2111.14551 [hep-th]].


\bibitem{Goswami:2021ksw}
K.~Goswami, K.~Narayan and H.~K.~Saini,
``Cosmologies, singularities and quantum extremal surfaces,''
JHEP \textbf{03}, 201 (2022)
doi:10.1007/JHEP03(2022)201
[arXiv:2111.14906 [hep-th]].

\bibitem{Bousso:2022gth}
R.~Bousso and E.~Wildenhain,
``Islands in closed and open universes,''
Phys. Rev. D \textbf{105}, no.8, 086012 (2022)
doi:10.1103/PhysRevD.105.086012
[arXiv:2202.05278 [hep-th]].

\bibitem{Moitra:2022glw}
U.~Moitra, S.~K.~Sake and S.~P.~Trivedi,
``Aspects of Jackiw-Teitelboim gravity in Anti-de Sitter and de Sitter spacetime,''
JHEP \textbf{06}, 138 (2022)
doi:10.1007/JHEP06(2022)138
[arXiv:2202.03130 [hep-th]].

\bibitem{Svesko:2022txo}
A.~Svesko, E.~Verheijden, E.~P.~Verlinde and M.~R.~Visser,
``Quasi-local energy and microcanonical entropy in two-dimensional nearly de Sitter gravity,''
JHEP \textbf{08}, 075 (2022)
doi:10.1007/JHEP08(2022)075
[arXiv:2203.00700 [hep-th]].

\bibitem{Ageev:2022hqc}
D.~S.~Ageev and I.~Y.~Aref'eva,
``Thermal density matrix breaks down the Page curve,''
[arXiv:2206.04094 [hep-th]].

\bibitem{Karch:2022rvr}
A.~Karch, H.~Sun and C.~F.~Uhlemann,
``Double holography in string theory,''
JHEP \textbf{10}, 012 (2022)
doi:10.1007/JHEP10(2022)012
[arXiv:2206.11292 [hep-th]].

\bibitem{Goswami:2022ylc}
K.~Goswami and K.~Narayan,
``Small Schwarzschild de Sitter black holes, quantum extremal surfaces and islands,''
JHEP \textbf{10}, 031 (2022)
doi:10.1007/JHEP10(2022)031
[arXiv:2207.10724 [hep-th]].

\bibitem{RoyChowdhury:2022awr}
A.~Roy Chowdhury, A.~Saha and S.~Gangopadhyay,
``Role of mutual information in the Page curve,''
Phys. Rev. D \textbf{106}, no.8, 086019 (2022)
[arXiv:2207.13029 [hep-th]].

\bibitem{Yu:2022xlh}
M.~H.~Yu and X.~H.~Ge,
``Entanglement islands in generalized two-dimensional dilaton black holes,''
Phys. Rev. D \textbf{107}, no.6, 066020 (2023)
[arXiv:2208.01943 [hep-th]].

\bibitem{Bousso:2022hlz}
R.~Bousso and G.~Penington,
``Entanglement wedges for gravitating regions,''
Phys. Rev. D \textbf{107}, no.8, 086002 (2023)
doi:10.1103/PhysRevD.107.086002
[arXiv:2208.04993 [hep-th]].

\bibitem{Hu:2022zgy}
P.~J.~Hu, D.~Li and R.~X.~Miao,
``Island on codimension-two branes in AdS/dCFT,''
JHEP \textbf{11}, 008 (2022)
doi:10.1007/JHEP11(2022)008
[arXiv:2208.11982 [hep-th]].

\bibitem{Ageev:2022qxv}
D.~S.~Ageev, I.~Y.~Aref'eva, A.~I.~Belokon, A.~V.~Ermakov, V.~V.~Pushkarev, T.~A.~Rusalev,
``Infrared regularization and finite size dynamics of entanglement entropy in Schwarzschild black hole,''
Phys. Rev. D \textbf{108}, no.4, 046005 (2023)
[arXiv:2209.00036 [hep-th]].

\bibitem{Chu:2022ieq}
C.~S.~Chu, R.~X.~Miao,
``Tunneling of Bell Particles, Page Curve and Black Hole Information,''
[arXiv:2209.03610 [hep-th]].

\bibitem{Yadav:2022jib}
G.~Yadav and N.~Joshi,
``Cosmological and black hole islands in multi-event horizon spacetimes,''
Phys. Rev. D \textbf{107}, no.2, 026009 (2023)
doi:10.1103/PhysRevD.107.026009
[arXiv:2210.00331 [hep-th]].

\bibitem{Aalsma:2022swk}
L.~Aalsma, S.~E.~Aguilar-Gutierrez and W.~Sybesma,
``An outsider\textquoteright{}s perspective on information recovery in de Sitter space,''
JHEP \textbf{01}, 129 (2023)
doi:10.1007/JHEP01(2023)129
[arXiv:2210.12176 [hep-th]].

\bibitem{Lu:2022tmt}
C.~Y.~Lu, M.~H.~Yu, X.~H.~Ge and L.~J.~Tian,
``Page curve and phase transition in deformed Jackiw\textendash{}Teitelboim gravity,''
Eur. Phys. J. C \textbf{83}, no.3, 215 (2023)
doi:10.1140/epjc/s10052-023-11358-5
[arXiv:2210.14750 [hep-th]].

\bibitem{Stepanenko:2022gwy}
D.~Stepanenko and I.~Volovich,
``Schwarzschild black holes, Islands and Virasoro algebra,''
Eur. Phys. J. Plus \textbf{138}, no.8, 688 (2023)
doi:10.1140/epjp/s13360-023-04342-1
[arXiv:2211.03153 [hep-th]].

\bibitem{Ben-Dayan:2022nmb}
I.~Ben-Dayan, M.~Hadad and E.~Wildenhain,
``Islands in the fluid: islands are common in cosmology,''
JHEP \textbf{03}, 077 (2023)
doi:10.1007/JHEP03(2023)077
[arXiv:2211.16600 [hep-th]].

\bibitem{Basu:2022crn}
D.~Basu, Q.~Wen and S.~Zhou,
``Entanglement Islands from Hilbert Space Reduction,''
[arXiv:2211.17004 [hep-th]].

\bibitem{Kudler-Flam:2022irq}
J.~Kudler-Flam and Y.~Kusuki,
``On quantum information before the Page time,''
JHEP \textbf{05}, 078 (2023)
doi:10.1007/JHEP05(2023)078
[arXiv:2212.06839 [hep-th]].

\bibitem{Emparan:2023dxm}
R.~Emparan, R.~Luna, R.~Suzuki, M.~Toma\v{s}evi\'c and B.~Way,
``Holographic duals of evaporating black holes,''
JHEP \textbf{05}, 182 (2023)
doi:10.1007/JHEP05(2023)182
[arXiv:2301.02587 [hep-th]].

\bibitem{Piao:2023vgm}
Y.~S.~Piao,
``Implication of the island rule for inflation and primordial perturbations,''
Phys. Rev. D \textbf{107}, no.12, 123509 (2023)
doi:10.1103/PhysRevD.107.123509
[arXiv:2301.07403 [hep-th]].

\bibitem{Guo:2023gfa}
C.~Z.~Guo, W.~C.~Gan and F.~W.~Shu,
``Page curves and entanglement islands for the step-function Vaidya model of evaporating black holes,''
JHEP \textbf{05}, 042 (2023)
doi:10.1007/JHEP05(2023)042
[arXiv:2302.02379 [hep-th]].

\bibitem{Parvizi:2023foz}
S.~Parvizi and M.~Shahbazi,
``Analogue gravity and the island prescription,''
Eur. Phys. J. C \textbf{83}, no.8, 705 (2023)
doi:10.1140/epjc/s10052-023-11874-4
[arXiv:2302.08742 [hep-th]].

\bibitem{Hung:2023mbw}
T.~N.~Hung and C.~H.~Nam,
``Compactified extra dimension and entanglement island as clues to quantum gravity,''
Eur. Phys. J. C \textbf{83}, no.6, 472 (2023)
doi:10.1140/epjc/s10052-023-11606-8
[arXiv:2303.00348 [hep-th]].

\bibitem{Wu:2023uyb}
C.~H.~Wu and J.~Xu,
``Islands in non-minimal dilaton gravity: exploring effective theories for black hole evaporation,''
JHEP \textbf{10}, 094 (2023)
doi:10.1007/JHEP10(2023)094
[arXiv:2303.03410 [hep-th]].

\bibitem{Cadoni:2023tse}
M.~Cadoni, M.~Oi and A.~P.~Sanna,
``Evaporation and information puzzle for 2D nonsingular asymptotically flat black holes,''
JHEP \textbf{06}, 211 (2023)
doi:10.1007/JHEP06(2023)211
[arXiv:2303.05557 [hep-th]].

\bibitem{RoyChowdhury:2023eol}
A.~Roy Chowdhury, A.~Saha and S.~Gangopadhyay,
``Mutual information of subsystems and the Page curve for the Schwarzschild\textendash{}de Sitter black hole,''
Phys. Rev. D \textbf{108}, no.2, 026003 (2023)
[arXiv:2303.14062 [hep-th]].

\bibitem{Ageev:2023mzu}
D.~S.~Ageev, I.~Y.~Aref'eva, A.~I.~Belokon, V.~V.~Pushkarev and T.~A.~Rusalev,
``Entanglement entropy in de Sitter: no pure states for conformal matter,''
[arXiv:2304.12351 [hep-th]].

\bibitem{Basu:2023wmv}
D.~Basu, J.~Lin, Y.~Lu and Q.~Wen,
``Ownerless island and partial entanglement entropy in island phases,''
[arXiv:2305.04259 [hep-th]].

\bibitem{Jeong:2023lkc}
H.~S.~Jeong, K.~Y.~Kim and Y.~W.~Sun,
``Entanglement entropy analysis of dyonic black holes using doubly holographic theory,''
Phys. Rev. D \textbf{108}, no.12, 126016 (2023)
doi:10.1103/PhysRevD.108.126016
[arXiv:2305.18122 [hep-th]].

\bibitem{Tong:2023nvi}
C.~W.~Tong, D.~H.~Du and J.~R.~Sun,
``Island of Reissner-Nordstr$\mathbf{\ddot{o}}$m anti-de Sitter black holes in the large $d$ limit,''
[arXiv:2306.06682 [hep-th]].

\bibitem{Yu:2023whl}
M.~H.~Yu, X.~H.~Ge and C.~Y.~Lu,
``Page Curves for Accelerating Black Holes,''
[arXiv:2306.11407 [hep-th]].

\bibitem{Chou:2023adi}
C.~J.~Chou, H.~B.~Lao and Y.~Yang,
``Page Curve of AdS-Vaidya Model for Evaporating Black Holes,''
[arXiv:2306.16744 [hep-th]].

\bibitem{Aguilar-Gutierrez:2023ymx}
S.~E.~Aguilar-Gutierrez, R.~Esp\'\i{}ndola and E.~K.~Morvan-Benhaim,
``A teleportation protocol in Schwarzschild-de Sitter space,''
[arXiv:2308.13516 [hep-th]].

\bibitem{Czech:2023rbh}
B.~Czech, S.~Shuai and H.~Tang,
``Information recovery in the Hayden-Preskill protocol,''
[arXiv:2310.16988 [hep-th]].

\bibitem{Franken:2023jas}
V.~Franken, H.~Partouche, F.~Rondeau and N.~Toumbas,
``Closed FRW holography: A time-dependent ER=EPR realization,''
[arXiv:2310.20652 [hep-th]].

\bibitem{Narayan:2015vda}
K.~Narayan,
``Extremal surfaces in de Sitter spacetime,''
Phys. Rev. D \textbf{91}, no.12, 126011 (2015)
doi:10.1103/PhysRevD.91.126011
[arXiv:1501.03019 [hep-th]].

\bibitem{Sato:2015tta}
Y.~Sato,
``Comments on Entanglement Entropy in the dS/CFT Correspondence,''
Phys. Rev. D \textbf{91}, no.8, 086009 (2015)
doi:10.1103/PhysRevD.91.086009
[arXiv:1501.04903 [hep-th]].

\bibitem{Narayan:2017xca}
K.~Narayan,
``On extremal surfaces and de Sitter entropy,''
Phys. Lett. B \textbf{779}, 214-222 (2018)
doi:10.1016/j.physletb.2018.02.010
[arXiv:1711.01107 [hep-th]].

\bibitem{Doi:2022iyj}
K.~Doi, J.~Harper, A.~Mollabashi, T.~Takayanagi and Y.~Taki,
``Pseudoentropy in dS/CFT and Timelike Entanglement Entropy,''
Phys. Rev. Lett. \textbf{130}, no.3, 031601 (2023)
doi:10.1103/PhysRevLett.130.031601
[arXiv:2210.09457 [hep-th]].

\bibitem{Narayan:2022afv}
K.~Narayan,
``de Sitter space, extremal surfaces, and time entanglement,''
Phys. Rev. D \textbf{107}, no.12, 126004 (2023)
doi:10.1103/PhysRevD.107.126004
[arXiv:2210.12963 [hep-th]].

\bibitem{Gibbons:1977mu} 
  G.~W.~Gibbons and S.~W.~Hawking,
  ``Cosmological Event Horizons, Thermodynamics, and Particle Creation,''
  Phys.\ Rev.\ D {\bf 15}, 2738 (1977).
  doi:10.1103/PhysRevD.15.2738

\bibitem{Ginsparg:1982rs} 
  P.~H.~Ginsparg and M.~J.~Perry,
  ``Semiclassical Perdurance of de Sitter Space,''
  Nucl.\ Phys.\ B {\bf 222}, 245 (1983).
  doi:10.1016/0550-3213(83)90636-3

\bibitem{Bousso:1996au} 
  R.~Bousso and S.~W.~Hawking,
  ``Pair creation of black holes during inflation,''
  Phys.\ Rev.\ D {\bf 54}, 6312 (1996)
  doi:10.1103/PhysRevD.54.6312
  [gr-qc/9606052].

\bibitem{Bousso:1997wi}
R.~Bousso and S.~W.~Hawking,
``(Anti)evaporation of Schwarzschild-de Sitter black holes,''
Phys. Rev. D \textbf{57}, 2436-2442 (1998)
doi:10.1103/PhysRevD.57.2436
[arXiv:hep-th/9709224 [hep-th]].

\bibitem{Nariai}
  H. Nariai,
  ``On some static solutions of Einstein’s gravitational field equations
  in a spherically symmetric case'',
  Sci. Rep. Tohoku Univ. Eighth Ser. 34, 1950.

\bibitem{Maldacena:2019cbz} 
  J.~Maldacena, G.~J.~Turiaci and Z.~Yang,
  ``Two dimensional Nearly de Sitter gravity,''
  arXiv:1904.01911 [hep-th].

\bibitem{Fernandes:2019ige}
K.~Fernandes, K.~S.~Kolekar, K.~Narayan and S.~Roy,
``Schwarzschild de Sitter and extremal surfaces,''
Eur. Phys. J. C \textbf{80}, no.9, 866 (2020)
doi:10.1140/epjc/s10052-020-08437-2
[arXiv:1910.11788 [hep-th]].

\bibitem{Shankaranarayanan:2003ya}
S.~Shankaranarayanan,
``Temperature and entropy of Schwarzschild-de Sitter space-time,''
Phys. Rev. D \textbf{67}, 084026 (2003)
doi:10.1103/PhysRevD.67.084026
[arXiv:gr-qc/0301090 [gr-qc]].

\bibitem{Guven:1990ubi}
J.~Guven and D.~N\'u\~nez,
``Schwarzschild-de Sitter space and its perturbations,''
Phys. Rev. D \textbf{42}, no.8, 2577-2584 (1990)
doi:10.1103/physrevd.42.2577

\bibitem{Strominger:1994tn}
A.~Strominger,
``Les Houches lectures on black holes,''
[arXiv:hep-th/9501071 [hep-th]].

\bibitem{Grumiller:2002nm}
D.~Grumiller, W.~Kummer and D.~V.~Vassilevich,
``Dilaton gravity in two-dimensions,''
Phys. Rept. \textbf{369}, 327-430 (2002)
doi:10.1016/S0370-1573(02)00267-3
[arXiv:hep-th/0204253 [hep-th]].

\bibitem{Mertens:2022irh}
T.~G.~Mertens and G.~J.~Turiaci,
``Solvable models of quantum black holes: a review on Jackiw\textendash{}Teitelboim gravity,''
Living Rev. Rel. \textbf{26}, no.1, 4 (2023)
doi:10.1007/s41114-023-00046-1
[arXiv:2210.10846 [hep-th]].

\bibitem{Narayan:2020pyj}
K.~Narayan,
``On aspects of two-dimensional dilaton gravity, dimensional reduction, and holography,''
Phys. Rev. D \textbf{104}, no.2, 026007 (2021)
doi:10.1103/PhysRevD.104.026007
[arXiv:2010.12955 [hep-th]].

\bibitem{Bhattacharya:2020qil}
R.~Bhattacharya, K.~Narayan and P.~Paul,
``Cosmological singularities and 2-dimensional dilaton gravity,''
JHEP \textbf{08}, 062 (2020)
doi:10.1007/JHEP08(2020)062
[arXiv:2006.09470 [hep-th]].

\bibitem{Calabrese:2004eu}
P.~Calabrese and J.~L.~Cardy,
``Entanglement entropy and quantum field theory,''
J. Stat. Mech. \textbf{0406}, P06002 (2004)
doi:10.1088/1742-5468/2004/06/P06002
[arXiv:hep-th/0405152 [hep-th]].

\bibitem{Calabrese:2009qy}
P.~Calabrese and J.~Cardy,
``Entanglement entropy and conformal field theory,''
J. Phys. A \textbf{42}, 504005 (2009)
doi:10.1088/1751-8113/42/50/504005
[arXiv:0905.4013 [cond-mat.stat-mech]].

\bibitem{Susskind:2007pv}
L.~Susskind,
``The Census taker's hat,''
[arXiv:0710.1129 [hep-th]].

\bibitem{Calabrese:2009ez}
P.~Calabrese, J.~Cardy, E.~Tonni,
``Entanglement entropy of two disjoint intervals in conformal field theory,''
J. Stat. Mech. \textbf{0911}, P11001 (2009)\! 
doi:10.1088/1742-5468/2009/11/P11001\!
[arXiv:0905.2069\,[hep-th]].

\bibitem{Calabrese:2010he}
P.~Calabrese, J.~Cardy, E.~Tonni,
``Entanglement entropy of two disjoint intervals in conformal field theory II,''
J. Stat. Mech. \textbf{1101}, P01021 (2011)\!
doi:10.1088/1742-5468/2011/01/P01021\!
[arXiv:1011.5482\,[hep-th]].

\bibitem{Headrick:2010zt} 
  M.~Headrick,
  ``Entanglement Renyi entropies in holographic theories,''
  Phys.\ Rev.\ D {\bf 82}, 126010 (2010)
  [arXiv:1006.0047 [hep-th]].

\bibitem{Pedraza:2021cvx}
J.~F.~Pedraza, A.~Svesko, W.~Sybesma and M.~R.~Visser,
``Semi-classical thermodynamics of quantum extremal surfaces in Jackiw-Teitelboim gravity,''
JHEP \textbf{12}, 134 (2021)
doi:10.1007/JHEP12(2021)134
[arXiv:2107.10358 [hep-th]].

\bibitem{Pedraza:2021ssc}
J.~F.~Pedraza, A.~Svesko, W.~Sybesma and M.~R.~Visser,
``Microcanonical action and the entropy of Hawking radiation,''
Phys. Rev. D \textbf{105}, no.12, 126010 (2022)
doi:10.1103/PhysRevD.105.126010
[arXiv:2111.06912 [hep-th]].

\bibitem{Morvan:2022ybp}
E.~K.~Morvan, J.~P.~van der Schaar and M.~R.~Visser,
``On the Euclidean action of de Sitter black holes and constrained instantons,''
SciPost Phys. \textbf{14}, no.2, 022 (2023)
doi:10.21468/SciPostPhys.14.2.022
[arXiv:2203.06155 [hep-th]].


} }
\end{thebibliography}
\end{document}